\documentclass[aps,prd,twocolumn,superscriptaddress,showpacs,preprintnumbers,floatfix]{revtex4}
\usepackage{bm}
\usepackage{graphicx}

\newcommand{\xvec}{\bm{x}}
\newcommand{\yvec}{\bm{y}}

\newcommand{\pvec}{\bm{p}}

\begin{document}
\preprint{DESY 11-035}
\title{Improved stochastic estimation of quark propagation with Laplacian
       Heaviside smearing in lattice QCD}
\author{C.~Morningstar}
\affiliation{Department of Physics, 
             Carnegie Mellon University, 
             Pittsburgh, PA 15213, USA}
\author{J.~Bulava}
\affiliation{NIC, DESY, 
             Platanenallee 6, D-15738, 
             Zeuthen, Germany}
\author{J.~Foley}
\affiliation{Dept.\ of Physics and Astronomy, 
             University of Utah, 
             Salt Lake City, UT 84112, USA}
\author{K.J.~Juge}
\affiliation{Dept.~of Physics, 
             University of the Pacific, 
             Stockton, CA 95211, USA}
\author{D.~Lenkner}
\affiliation{Department of Physics, 
             Carnegie Mellon University, 
             Pittsburgh, PA 15213, USA}
\author{M.~Peardon}
\affiliation{School of Mathematics, 
             Trinity College, 
             Dublin 2, Ireland}
\author{C.H.~Wong}
\affiliation{Department of Physics, 
             Carnegie Mellon University, 
             Pittsburgh, PA 15213, USA}
\date{April 19, 2011}

\begin{abstract}
A new method of stochastically estimating the low-lying effects of quark
propagation is proposed which allows accurate determinations of temporal
correlations of single-hadron and multi-hadron operators in lattice QCD.
The method is well suited for calculations in large volumes. 
Contributions involving quark propagation connecting hadron sink 
operators at the same final time can be handled in a straightforward manner, 
even for a large number of final time slices.  The method exploits Laplacian 
Heaviside (LapH) smearing.  $Z_N$ noise is introduced in a novel way, and 
variance reduction is achieved using judiciously-chosen noise dilution 
projectors.  The method is tested using isoscalar mesons in the scalar, 
pseudoscalar, and vector channels, and using the two-pion system of total
isospin $I=0,1,2$ on large anisotropic $24^3\times 128$ lattices with 
spatial spacing $a_s\sim 0.12$~fm and temporal spacing $a_t\sim 0.034$~fm 
for pion masses $m_\pi\approx 390$ and 240~MeV.
\end{abstract}
\pacs{12.38.Gc, 11.15.Ha, 12.39.Mk}
\maketitle

\section{Introduction}
Recent discoveries of new hadronic resonances have generated much
excitement in the field of hadron spectroscopy.  The current surge in 
experimental activity underlines the need for a better understanding of 
excited hadronic states from the theory of quantum chromodynamics (QCD). 
Presently, Markov-chain Monte Carlo estimates of QCD path integrals 
defined on a space-time lattice offer the best way to make progress 
in this regard.

Calculating the mass spectrum of excited-state hadron resonances is a key 
goal in lattice QCD.  However, such calculations are very challenging.  
Computational limitations cause simulations to be done with quark masses 
that are unphysically large, leading to pion masses that are heavier 
than observed and introducing systematic errors in all other hadron energies.  
The use of carefully designed quantum field operators is crucial 
for accurate determinations of low-lying energies. To study a particular state 
of interest, the energies of all states lying below that state must first be 
extracted, and as the pion gets lighter in lattice QCD simulations, more and 
more multi-hadron states lie below the masses of the excited resonances.  The 
evaluation of correlations involving multi-hadron operators contains new 
challenges since not only must initial to final time quark propagation be 
included, but also final to final time quark propagation must be incorporated.  
The masses and widths of resonances (unstable hadrons) cannot be calculated 
directly in finite-volume Monte Carlo computations, but must be deduced from 
the discrete spectrum of finite-volume stationary states for a range of box 
sizes\cite{DeWitt:1956be,Wiese:1988qy,Luscher:1991cf,Rummukainen:1995vs}.  

Our approach to constructing hadron operators appropriate
for such calculations was outlined in Refs.~\cite{baryons2005A,baryons2005B}.
Our first study of the nucleon and $\Delta$ excitations in the quenched
approximation was presented in Ref.~\cite{baryon2007}, and nucleon results 
for two flavors of dynamical quarks appeared in Ref.~\cite{nucleon2009}.
A survey of excited-state energies in small volume for the isovector mesons
and kaons using $N_f=2+1$ dynamical quarks was given in 
Ref.~\cite{Morningstar:2010ae}, along with results for the $\Lambda, \Sigma, 
\Xi$ baryons.  Other recent progress in calculating 
excited-state energies in lattice
QCD can be found in Refs.~\cite{edwards2009,
Mahbub:2009nr,Mahbub:2010jz,Engel:2010my,Bulava:2010yg,Peardon:2011tt}.
All of our results to date have been
achieved in small volume with pions having masses comparable to or heavier
than about 390~MeV.  Our goal now is to extend our efforts into larger
volumes and using lighter pions.  To do this, the issue of multi-hadron
states must be addressed.

In this work, we focus on the problem of incorporating multi-hadron operators 
into finite-volume excited-state spectrum calculations in lattice QCD.  
To compute the finite-volume stationary-state energies of QCD, one must
first evaluate a matrix of temporal correlations
$C_{ij}(t_F-t_0)=\langle 0\vert\,T\, O_i(t_F)\,\overline{O}_j(t_0)\,\vert 0\rangle$,
where $T$ denotes time-ordering, the source operators $\overline{O}_j(t_0)$ create 
the states of interest at an initial time $t_0$, and the sink operators $O_i(t_F)$ 
annihilate the states of interest at a later time $t_F$.  
The correlation functions $C_{ij}(t)$ can be expressed in terms of ``path" 
integrals over quark $\overline{\psi},\psi$ fields and gluon $U$ fields
involving the QCD action having the form
\begin{equation}
   S[\overline{\psi},\psi,U] = \overline{\psi} K[U]\psi + S_G[U],
\end{equation}
where $K[U]$ is known as the Dirac matrix and $S_G[U]$ is the gauge-field
action.  Integration over the Grassmann-valued quark fields introduces
a $\det K$ and factors of $K^{-1}$ in the remaining integrals over
the gluon $U$ field, and when formulated on a Euclidean space-time lattice, such 
path integrals can be estimated using the Monte Carlo method with Markov-chain 
importance sampling.  Incorporating the $\det K$ in the Monte Carlo
updating and evaluating the elements of $K^{-1}$ (the quark propagators) are 
the most computationally demanding parts of the calculations. 

Once estimates of a Hermitian matrix of temporal correlation functions 
$C_{ij}(t)$ are obtained, several procedures for extracting the lowest 
stationary-state energies $E_0,E_1,E_2,\dots$ in any given symmetry channel 
are available\cite{Michael:1985ne,Luscher:1990ck}.  For example, let 
$\lambda_n(t,t_0)$ denote the eigenvalues of the Hermitian matrix 
$C(t_0)^{-1/2}\,C(t)\,C(t_0)^{-1/2}$, where $t_0$ is some fixed reference time
(typically small) and the eigenvalues, also known as the {\em principal} 
correlation functions, are ordered such that $\lambda_0\geq\lambda_1\geq\cdots$ 
as $t$ becomes large.  Then one can show that
\begin{eqnarray}
 \lim_{t\rightarrow\infty}\lambda_n(t,t_0) &=& e^{-E_n (t-t_0)}.
  \label{eq:corrmat}
\end{eqnarray}
Determinations of the principal correlators $\lambda_n(t,t_0)$ for sufficiently 
large temporal separations $t$ yield the desired energies $E_n$.

The rows and columns of the gauge-covariant Dirac matrix $K[U]$ can be viewed
as compound indices which incorporate the lattice space-time site indices and the 
quark color, flavor, and spin indices.  Hence, $K$ is a very large matrix, and 
determining and storing all of the elements of $K^{-1}$ is not possible.  Symmetries
are used to eliminate the need to compute all $K^{-1}$ elements.  Computations are 
usually arranged such that the linear system of equations $Kx=y$ needs to be solved
for only a manageable number of source vectors $y$.  For temporal correlations of
single-hadron operators (excluding isoscalar mesons), invariance under all 
spatial and temporal translations dramatically reduces the number of $K^{-1}$
elements required.  In such cases, the hadron creation operator needs to be 
considered only on one initial time slice and only at a single spatial site, 
yielding the so-called point-to-all method.  A handful of points can be used
to increase statistics.

\begin{figure}
\includegraphics[width=3.3in,bb=0 30 743 324]{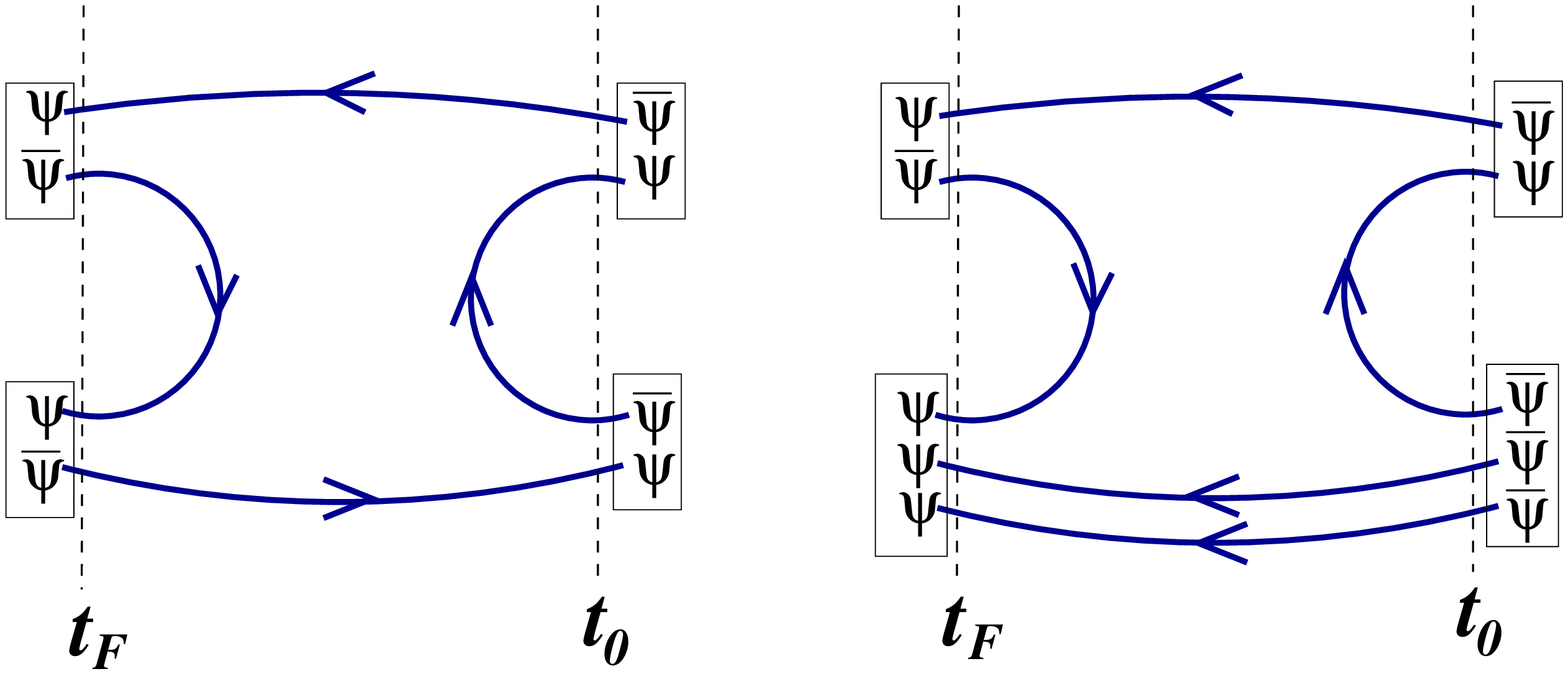}
\caption[figmulticorr]{
Examples of quark-line diagrams in multi-hadron correlators involving 
the $\overline{\psi}$ field on the later time $t_F$ connecting to a $\psi$
field also on the later time slice $t_F$. The initial time
is denoted by $t_0$. (Left) A two-meson 
correlator. (Right) The correlator of a baryon-meson system.
\label{fig:multicorr}}
\end{figure}

To study a particular eigenstate, the procedure by which energies
are extracted from Monte Carlo estimates of temporal correlation functions
using Eq.~(\ref{eq:corrmat}) requires that all eigenstates lying below 
the state of interest must first be extracted. As the pion gets lighter in
lattice QCD simulations, more and 
more \textit{multi-hadron} states will lie below the excited resonances, and
multi-hadron operators will be needed to accurately compute the energies of such
states.  For example, a good baryon-meson sink operator which annihilates a total
zero momentum is typically a superposition of terms having the form
\[
 B(-\pvec,t)M(\pvec,t)=\frac{1}{V^2}\sum_{\xvec,\yvec}\varphi_B(\xvec,t)
\varphi_M(\yvec,t)e^{i\pvec\cdot(\xvec-\yvec)},
\]
where $V$ is the spatial volume of the lattice, $2\pvec$ is the relative
momentum, and $\varphi_B(\xvec,t)$ and $\varphi_M(\yvec,t)$ are appropriate 
localized interpolating fields for a baryon and a meson, respectively.  
In the evaluation of the temporal correlations of such a multi-hadron operator, 
it is not possible to completely remove all summations over the spatial sites on 
the source time slice using
translation invariance.  Hence, the need for estimates of the quark propagators 
$K^{-1}$ from all spatial sites on a time slice to all spatial sites on another 
time slice cannot be sidestepped.  Some correlators involve diagrams with
quark lines originating at the sink time $t_F$ and terminating at the
same sink time $t_F$ (see Fig.~\ref{fig:multicorr}), so quark propagators 
involving a large number of quark-line starting times must also be handled.

Finding better ways to incorporate the low-lying effects of such slice-to-slice
quark propagation for large numbers of quark source times is crucial to
the success of our excited-state hadron spectrum project at lighter pion masses.
A new method, known as distillation\cite{distillation2009}, uses a novel
quark-field smearing procedure that facilitates exact treatment of
slice-to-slice quark propagation.  Although distillation was found to work
well, calculations with that method are costly, making it feasible only
on small lattices.  Here, we propose to combine the quark-field smearing used
in Ref.~\cite{distillation2009} with a new stochastic approach to 
estimating the quark propagators, resulting in a much more efficient
treatment suitable for large volumes.  Describing and testing this new method
is the aim of this work.  This method was briefly introduced with 
preliminary testing in Refs.~\cite{Morningstar:2010ae,Foley:2010vv,Bulava:2010em}.

The remainder of this paper is organized as follows.  The stochastic LapH 
method is described in Sec.~\ref{sec:method}.  Laplacian Heaviside quark-field
smearing is reviewed, and our new stochastic treatment of quark propagation
is detailed.  The method involves Monte Carlo estimation of quark propagation
using $Z_N$ noise in the LapH subspace with variance reduction achieved
through the introduction of suitable noise dilution projectors.  The new method
is compared to an earlier procedure which uses noise introduced on the 
space-time lattice itself.  The number of inversions of the Dirac
matrix needed in the new method is demonstrated to be insensitive
to the volume of the lattice.   Details on how the temporal correlations
of hadron operators are evaluated using the stochastic LapH method are
then presented in Sec.~\ref{sec:corr}.  Full source-sink factorization is seen
to be another advantageous feature of the method, especially for 
computations of correlation matrices involving large sets of hadron operators. 
Various implementation details are given in Sec.~\ref{sec:implement}.   
Applications of the method to the isoscalar mesons in the scalar, pseudoscalar,
and vector channels and to the 
two-pion system of total isospin $I=0,1,2$ using anisotropic $24^3\times 128$
lattices with pion masses $m_\pi\approx 390$ and 240~MeV are then presented 
in Sec.~\ref{sec:testing}. Conclusions are summarized in Sec.~\ref{sec:conclude}. 

\section{Description of the method}
\label{sec:method}

The use of smeared fields is crucial for successfully extracting the 
spectrum of QCD in our Monte Carlo computations.  Hadron operators constructed
out of smeared fields have dramatically reduced mixings with the high-frequency
modes of the theory that obscure the low-lying eigenstates of interest.
Our operators are constructed using spatially-smoothed link variables 
$\widetilde{U}_j(x)$ and spatially-smeared quark fields $\widetilde{\psi}(x)$.  

The spatial links are smeared using the
stout-link procedure described in Ref.~\cite{Morningstar:2003gk}.
Note that only spatial staples are used in the link smoothening; no temporal
staples are used, and the temporal link variables are not smeared.

The quark field for each quark flavor is smeared using
\begin{equation}
\widetilde{\psi}_{a\alpha}(x) =
   {\cal S}_{ab}(x,y)\ \psi_{b\alpha}(y),
\end{equation}
where $x,y$ are lattice sites, $a,b$ are color indices, $\alpha$ is a 
Dirac spin component, and the smearing kernel ${\cal S}$ is such that the 
smeared field behaves in exactly the same way as the original field under all
time-independent symmetry transformations on a cubic lattice.  For extracting
energies from temporal correlations, it is important that only spatial smearing
is used.  In other words, the smearing kernel is diagonal in time: 
${\cal S}_{ab}(x,y) \propto \delta_{x_4 y_4}$.  In addition, our smearing 
kernel is independent of spin.

We use the new Laplacian Heaviside (LapH) quark-field smearing scheme
which has been described in Ref.~\cite{distillation2009} 
and is defined by
\begin{equation}
{\cal S} = 
 \Theta\left(\sigma_s^2+\widetilde{\Delta}\right),
\end{equation}
where $\widetilde{\Delta}$ is the three-dimensional gauge-covariant Laplacian
defined in terms of the stout-smeared gauge field $\widetilde{U}$, and $\sigma_s$
is the smearing cutoff parameter.  The Laplacian matrix is given by
\begin{eqnarray}
   &&\widetilde{\Delta}^{ab}(x,y;U) = \sum_{k=1}^3 \Bigl\{
   \widetilde{U}_k^{ab}(x)\delta(y,x+\hat{k})  \nonumber\\
   &&\qquad + \widetilde{U}^{ba}_k(y)^\ast\delta(y,x-\hat{k})
   - 2\delta(x,y)\delta^{ab}\Bigr\},        
\end{eqnarray}
where $x,y$ are lattice sites, and $a,b$ are 
color indices.  This is a Hermitian matrix which is block-diagonal in time.
It is important to use the stout-smeared gauge links when smearing the quark
field since doing so dramatically reduces the statistical errors in the 
correlations of the hadron operators we use which involve 
covariantly-displaced quark fields\cite{nucleon2009}. The gauge-covariant 
Laplacian operator is
ideal for smearing the quark field since it is one of the simplest operators
that locally averages the field in such a way that all relevant symmetry
transformation properties of the original field are preserved.  

\begin{figure}[t]
\begin{center}
\includegraphics[width=3.2in,bb=19 34 571 552]{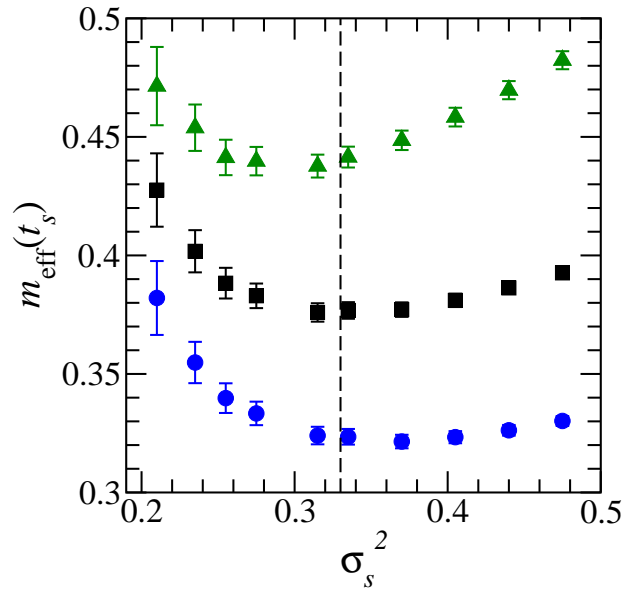}
\end{center}
\caption[figsigmachoose]{
 The effective masses for temporal separation $t_s=1$ for three
 representative nucleon operators against the LapH smearing cutoff
 $\sigma_s^2$. Results were obtained using $N_f=2+1$ configurations 
 on a $16^3\times 128$ anisotropic lattice with spacing $a_s\sim 0.12$~fm 
 for stout-link smearing with $n_\xi=10$ iterations and staple
 weight $\xi=0.1$.  The circles show
 results (shifted downward by 0.04) for a single-site operator.
 The squares correspond to a singly-displaced nucleon operator, and
 the triangles are the results (shifted upward by 0.04) for a
 triply-displaced-T operator.  The value $\sigma_s^2\approx 0.33$
 is observed to be a good choice.
\label{fig:sigmachoose}}
\end{figure}

Let $V_\Delta$ denote the unitary matrix whose columns are the eigenvectors
of $\widetilde{\Delta}$, and let $\Lambda_\Delta$ denote a diagonal matrix
whose elements are the eigenvalues of $\widetilde{\Delta}$ such that
\begin{equation}
    \widetilde{\Delta}=V_\Delta\ \Lambda_\Delta\ V_\Delta^\dagger.
\end{equation}
The LapH smearing matrix is then given by
\begin{equation}
    {\cal S} = V_\Delta\ \Theta\left(\sigma_s^2+\Lambda_\Delta\right)
   \ V_\Delta^\dagger. \label{eq:smearA}
\end{equation}
All of the eigenvalues in $\Lambda_\Delta$ are real and less than zero.
Hence, the matrix $\Theta(\sigma_s^2+\Lambda_\Delta)$ has unit entries
on those diagonal elements corresponding to eigenvalues whose magnitudes
are less than $\sigma_s^2$ and zero entries for all other elements.
Given that $\widetilde{\Delta}$ is block-diagonal in time, each eigenvector
has nonzero elements only on one time slice, so we can associate
any given eigenpair with that particular time.  The eigenvalues of 
$\widetilde{\Delta}$ always occur such that approximately
$N_v$ eigenvalues have magnitude smaller than $\sigma_s^2$ on each time
slice.  We have observed that the number of eigenvalues on each time slice
that survive the Heaviside function varies from time to time by only one
or two in cases where $N_v$ exceeds sixty or more.  Hence, the Heaviside
smearing matrix is well approximated by fixing $N_v$ to the same value
on each time slice and for each gauge configuration.

Let $V_s$ denote the matrix whose columns are in one-to-one correspondence
with the eigenvectors associated with the $N_v$ lowest-lying eigenvalues of 
$-\widetilde{\Delta}$ on each time slice.  Then our LapH smearing matrix 
is well approximated by the Hermitian matrix
\begin{equation}
   {\cal S}=V_s\ V_s^\dagger.
\end{equation}
This is the actual smearing matrix used in our calculations.  Note that
on a lattice having $N_t$ time slices and $N_s$ sites in each of the
spatial directions, the matrix $V_s$ has $N_v N_t$ columns and 
$N_t N_s^3 N_c$ rows, where $N_c=3$ is the number of quark colors.
The $N_v N_t$ eigenvectors that form the smearing matrix span the
so-called LapH subspace.  

To set the parameter $\sigma_s$, and hence $N_v$,
several small simulations were done varying this parameter while computing the 
effective masses for a handful of simple meson and baryon operators.  We chose
the value of $\sigma_s$ that minimized the effective masses at a chosen early
time separation $t_s$.  The effective masses for $t_s=1$ for three representative
nucleon operators are shown in Fig.~\ref{fig:sigmachoose} against
values of $\sigma_s$.  A single-site nucleon operator in which all three quark
fields are taken at the same site is shown, as well as a singly-displaced
nucleon operator in which one of the quarks is displaced away from the other,
and a triply-displaced-T operator in which all three quarks are displaced
from the others in a T configuration.  The value $\sigma_s^2\approx 0.33$ was 
chosen.  This value is insensitive to which time interval is used as long 
as $t_s$ is small enough such that contributions from excited states have 
not decayed away. It is also insensitive to the choice of hadron
operator used and the quark mass.  
We expect $\sigma_s$ to change little with the lattice spacing.

Evaluating the temporal correlations of our hadron operators
requires combining matrix elements associated with various quark 
lines ${\cal Q}$.  Since we construct our hadron operators out of
covariantly-displaced, smeared quark fields, each and every quark line 
in our computation involves the following product of matrices:
\begin{equation}
 {\cal Q} = D^{(j)}{\cal S}\Omega^{-1}{\cal S}D^{(k)\dagger},
\end{equation}
where $\Omega=\gamma_4 K$ and
$D^{(i)}$ is a gauge-covariant displacement of type $i$.  The
displacement type can be trivial (no displacement), a displacement
in a given single spatial direction on the lattice by some number of links 
(typically three), or a combination of two or more spatial lattice directions.
The use of $\Omega=\gamma_4 K$ is convenient for ensuring baryon correlation
matrices that are Hermitian.

\begin{figure}[t]
\begin{center}
\includegraphics[width=3.2in,bb=0 81 537 523]{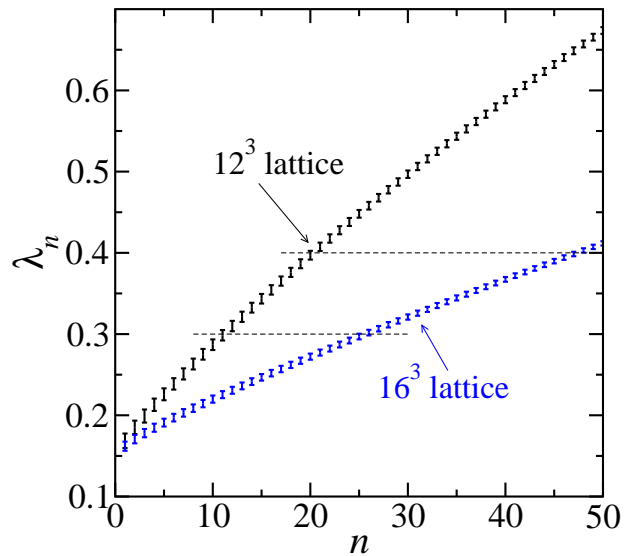}
\end{center}
\caption[figlapheigs1]{
 The effect of the spatial lattice volume on the eigenvalues of the
 gauge-covariant Laplacian operator. $\lambda_n$ is the $n$-th lowest eigenvalue 
 of $-\widetilde{\Delta}$ on a given time slice.  The error bars show
 the variation over different time slices and over a set of $N_f=2+1$ 
 configurations.  The lattice spacings $a_s$ are both near $0.12$~fm, and 
 the pion masses are both near 0.70~GeV.  Link smearing with $n_\xi=10$ iterations
 and staple weight $\xi=0.1$ was used.  For the $12^3$ lattice, there are nine
 eigenvalues between
 0.3 and 0.4, whereas for the $16^3$ lattice, there are 22 eigenvalues between 
 0.3 and 0.4, demonstrating that the density of eigenvalues is proportional
 to the spatial volume of the lattice (at sufficiently high values).
 The lowest-lying modes do not change very much with the lattice volume.
\label{fig:lapheigs1}}
\end{figure}

An exact treatment of such a quark line is best accomplished by writing
\begin{equation}
 {\cal Q} = D^{(j)} V_s\ (V_s^\dagger \Omega^{-1} V_s)
\ V_s^\dagger D^{(k)\dagger},
\end{equation}
then one needs to compute and store only the elements of the much 
smaller matrix $V_s^\dagger \Omega^{-1} V_s$
instead of computing and storing a very large number of $\Omega^{-1}$
elements.  Let $N_d=4$ denote the number
of Dirac spin components, and define
$y^{(i,\alpha)}_{c\beta}(x)=V_s(c,x;\ i)\ \delta_{\alpha\beta}$,
where $\alpha,\beta$ are spin indices, $c$ indicates color, $x$ is
a lattice site, and $i$ refers to the column of $V_s$ which is the $i$-th
eigenvector of the Laplacian. Then solving the linear system 
$\Omega x = y^{(i,\alpha)}$ for $x$ and all $i,\alpha$ by standard methods 
yields $\Omega^{-1} V_s^{(i)}$.  Hence, $N_v N_t N_d$
such inversions are required in order to obtain the full matrix
$V_s^\dagger \Omega^{-1} V_s$ for each quark mass and each gauge configuration
in the Monte Carlo ensemble.  If only one source time slice is used in the
hadron correlators, then $N_v N_d$ inversions are required per quark mass
per gauge configuration.  Once multi-hadron operators are included, however,
sink-to-sink quark lines are needed, so $N_v N_{\rm snk} N_d$ inversions
must be done, where $N_{\rm snk}$ is the number of sink times.  Generally,
a handful of hadron source times are used to improve statistics, so
upon including multi-hadron operators, one finds that the
number of inversions needed in practice ends up near $N_v N_t N_d$.

\begin{figure}[t]
\begin{center}
\includegraphics[width=3.2in,bb=0 81 537 523]{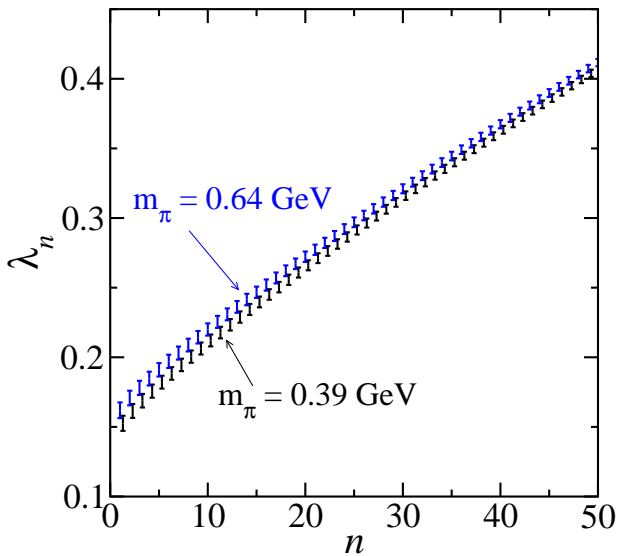}
\end{center}
\caption[figlapheigs2]{
 The small effect of the light-quark mass on the eigenvalues of the
 gauge-covariant Laplacian operator.  $\lambda_n$ is the $n$-th lowest 
 eigenvalue of $-\widetilde{\Delta}$ on a given time slice.  The error 
 bars show the variation over different time slices and over a set of 
 $N_f=2+1$ configurations on a $16^3\times 128$ anisotropic lattice with
 $a_s\sim 0.12$~fm for link smearing $n_\xi=10$ and $\xi=0.1$.  
\label{fig:lapheigs2}}
\end{figure}

Solving the linear systems $\Omega x=y$ is a major component of the computational
cost of evaluating the hadron correlators once a Monte Carlo ensemble is 
generated.  It turns out that the number $N_v$ of required eigenvectors on
each time slice rises in direct proportion to the spatial volume of
the lattice, as shown in Fig.~\ref{fig:lapheigs1}.  The number of
eigenvectors is also fairly insensitive to the light quark mass, as shown
in Fig.~\ref{fig:lapheigs2}.  Initial calculations on $16^3$ lattices with
spatial spacing $a_s\approx 0.12$~fm showed that $N_v=32$ worked well. On $20^3$
lattices, $N_v=64$ was needed, and for the $24^3\times 128$ lattices used
in this study, we found that $N_v=112$ levels were below the $\sigma_s^2$ cutoff. 
We have generated gauge configurations on $32^3\times 256$ anisotropic 
lattices.  On these lattices, we find that $N_v=264$, so the number of inversions
needed becomes $N_v N_t N_d > 270,000$ for each configuration and each quark mass, 
which is far too large to be feasible with current computing resources.

Fortunately, an exact treatment of the quark lines is not needed.  In fact,
we have found that exact treatment of the quark lines is very wasteful.
Given our use of the Monte Carlo method to evaluate the path integrals
over the gauge link variables, the statistical errors in our estimates of
the hadron correlators are ultimately limited by the statistical fluctuations
arising from the gauge-field sampling.  Thus, we only need to estimate the
quark lines to an accuracy comparable to the gauge noise from the Monte
Carlo method.  Such estimates can be obtained with far fewer inversions
than required by an exact treatment of the quark lines.

Random noise vectors $\eta$ which satisfy
$E(\eta_i)=0$ and $E(\eta_i\eta_j^\ast)=\delta_{ij}$, where $E()$ denotes
an expected value as defined in probability theory, are useful for 
stochastically estimating the inverse of a large matrix $\Omega$ 
as follows.  Assume that for each of $N_R$ 
noise vectors, we can solve the following
linear system of equations: $\Omega X^{r}=\eta^{r}$ for $X^{r}$, where
$r$ labels the noise vectors $r=1,2,\cdots,N_R$.
Then $X^{r}=\Omega^{-1}\eta^{r}$, and $E( X_i \eta_j^\ast ) = \Omega^{-1}_{ij}$
so that a Monte Carlo estimate of $\Omega_{ij}^{-1}$ is given by
$ \Omega_{ij}^{-1} \approx N_R^{-1}\sum_{r=1}^{N_R} X_i^{r}\eta_j^{r\ast}.$
Unfortunately, this equation usually produces stochastic estimates with 
variances which are much too large to be useful.  Variance reduction is
done by \textit{diluting} the noise 
vectors\cite{Bernardson:1993yg,Wilcox:1999ab,Foley:2005ac}.
A given dilution scheme can be viewed as the application of a complete
set of projection operators $P^{(b)}$. Define
$ \eta^{r[b]}=P^{(b)}\eta^{r} ,$
and define $X^{r[b]}$ as the solution of
$ \Omega X^{r[b]}=\eta^{r[b]},$
then a much better Monte Carlo estimate of $\Omega_{ij}^{-1}$ is
\begin{equation}
   \Omega_{ij}^{-1}\approx \frac{1}{N_R}
 \sum_{r=1}^{N_R} \sum_b X^{r[b]}_i\eta^{r[b]\ast}_j.
\label{eq:diluted}
\end{equation}
The dilution projections ensure \textit{exact zeros} for many
of the $E(\eta_i\eta_j^\ast)$ elements instead of estimates that are only 
statistically zero, resulting in a dramatic reduction
in the variance of the $\Omega^{-1}$ estimates.  The use of $Z_N$ noise ensures
zero variance in our estimates of the diagonal elements $E(\eta_i\eta_i^\ast)$. 
The effectiveness of the variance reduction depends on the projectors chosen.  

Earlier stochastic methods\cite{Edwards:2007en,Bulava:2008qx}
introduced noise in the full 
spin-color-space-time vector space, that is, on the entire lattice itself. 
However, since we intend to use Laplacian 
Heaviside quark-field smearing, an alternative is possible: noise vectors 
$\rho$ can be introduced \textit{only in the LapH subspace}.  The noise
vectors $\rho$ now have spin, time, and Laplacian eigenmode number
as their indices.  Color and space indices get replaced by Laplacian
eigenmode number.  Again, each component of $\rho$ is a random
$Z_N$ variable so that $E(\rho)=0$ and $E(\rho\rho^\dagger)=I_d$,
where $I_d$ is the identity matrix.
Dilution projectors $P^{(b)}$ are now matrices in the LapH subspace. 
In the stochastic LapH method, a quark line on a gauge configuration
is evaluated as follows:
\begin{eqnarray}
 {\cal Q}  &=&   D^{(j)} {\cal S} \Omega^{-1} {\cal S} D^{(k)\dagger},\nonumber\\
  &=&   D^{(j)} {\cal S} \Omega^{-1} V_s V_s^\dagger D^{(k)\dagger},\nonumber\\
  &=& \textstyle\sum_b  D^{(j)} {\cal S} \Omega^{-1} V_s P^{(b)}P^{(b)\dagger} 
 V_s^\dagger D^{(k)\dagger},\nonumber\\
&=& \textstyle\sum_b D^{(j)} {\cal S} \Omega^{-1} V_sP^{(b)}E(\rho\rho^\dagger)
  P^{(b)\dagger}  V_s^\dagger D^{(k)\dagger},\nonumber\\
 &=& \textstyle\sum_b E\Bigl( \! D^{(j)} {\cal S} \Omega^{-1} V_sP^{(b)}\rho
     \, (D^{(k)} V_s P^{(b)}  \rho)^\dagger \!\Bigr).
\end{eqnarray}
Displaced-smeared-diluted quark source 
and quark sink vectors can be defined by
\begin{eqnarray}
 \varrho^{[b]}(\rho) &=&  D^{(j)} V_s P^{(b)}\rho,
\label{eq:varrhodef}\\
 \varphi^{[b]}(\rho) &=&  D^{(j)} {\cal S} \Omega^{-1}\ V_s P^{(b)}\rho, 
\label{eq:varphidef}
\end{eqnarray}
and each quark line on a given gauge configuration can be estimated using
\begin{equation}
 {\cal Q}_{uv}^{(AB)} \approx \frac{1}{N_R}\delta_{AB}\sum_{r=1}^{N_R}\sum_b  
  \varphi^{[b]}_u(\rho^r)
  \  \varrho^{[b]}_v(\rho^r)^\ast,
\label{eq:laphprop}
\end{equation}
where the subscripts $u,v$ are compound indices combining space, time, color,
spin, and quark displacement type, $B$ is the flavor of the source field and
$A$ is the flavor of the sink field.  The above quark line estimate has the 
form of an outer product expansion.  Such estimates are frequently used in the 
compression of digital images, so the stochastic LapH estimate can be viewed 
as a lossy compression of the quark propagation.

Occasionally, it is useful to estimate a quark line using
$\gamma_5$-Hermiticity to switch the source and sink.  Using
$K^\dagger=\gamma_5 K\gamma_5$, it is straightforward to see that
another way to estimate a quark line is using
\begin{equation}
 {\cal Q}^{(AB)}_{uv} \approx \frac{1}{N_R}\delta_{AB}\sum_{r=1}^{N_R}\sum_b  
   \overline{\varrho}^{[b]}_u(\rho^r)
  \  \overline{\varphi}^{[b]}_v(\rho^r)^\ast,
\label{eq:laphpropg5}
\end{equation}
defining 
\begin{equation}
 \overline{\varrho}(\rho) = -\gamma_5\gamma_4\varrho(\rho),\qquad
 \overline{\varphi}(\rho) = \gamma_5\gamma_4 \varphi(\rho).
\end{equation}

Eqs.~(\ref{eq:laphprop}) and (\ref{eq:laphpropg5}) are meant to be used inside 
Monte Carlo estimates of path integrals over the gauge link variables. To 
simplify matters, the Monte Carlo within a Monte
Carlo computation can be combined into a single larger Monte Carlo calculation
over both gauge link variables and quark line noises, effectively setting 
$N_R=1$ for each gauge configuration.  However, each quark line in a hadron
correlator needs an independent noise to ensure unbiased estimation.  For example,
a baryon correlator requires at least three noises per gauge configuration.
Once inversions are done for a handful of such noise vectors for a given
configuration, noise permutations can be used to increase statistics.

The dilution projectors we use are products of time dilution, spin dilution, 
and LapH eigenvector dilution projectors.  The full projector index 
$b=(b_T,b_S,b_L)$ is a triplet of indices, where $b_T$ is the time projector 
index, $b_S$ is the spin projector index, and $b_L$ is the LapH eigenvector 
projector index. Our noise-dilution projectors have the form
\begin{equation}
 P^{(b)}_{t\alpha n;\ t^\prime\alpha^\prime n^\prime}
 = P^{(b_T)}_{t;t^\prime}\ P^{(b_S)}_{\alpha;\alpha^\prime}
 \ P^{(b_L)}_{n;n^\prime},
\end{equation}
where $t,t^\prime$ refer to time slices, $\alpha,\alpha^\prime$ are
Dirac spin indices, and $n,n^\prime$ are LapH eigenvector indices.
For each type (time, spin, LapH eigenvector) of dilution, we
studied four different dilution schemes.  Let $N$ denote the dimension
of the space of the dilution type of interest.  For time dilution, $N=N_t$
is the number of time slices on the lattice.  For spin dilution, $N=4$ is
the number of Dirac spin components.  For LapH eigenvector dilution, $N=N_v$
is the number of eigenvectors retained on each time slice.  The four 
schemes we studied are defined below:
\[ \begin{array}{lll}
P^{(b)}_{ij} = \delta_{ij},               & b=0,
   & \mbox{(no dilution)} \\
P^{(b)}_{ij} = \delta_{ij}\ \delta_{bi},  & b=0,\dots,N-1 
   & \mbox{(full dilution)}\\
P^{(b)}_{ij} = \delta_{ij}\ \delta_{b,\, \lfloor Ji/N\rfloor} 
   & b=0,\dots,J-1, & \mbox{(block-$J$)}\\
P^{(b)}_{ij} = \delta_{ij}\ \delta_{b,\, i\bmod J} & b=0,\dots,J-1, 
   & \mbox{(interlace-$J$)}
\end{array}\]
where $i,j=0,\dots,N-1$, and we assume $N/J$ is an integer.  Note that
each projector is a diagonal matrix with some or all of the diagonal elements
set to unity and all other elements vanishing.  We use a triplet
(T, S, L) to specify a given dilution scheme, where ``T" denote time,
``S" denotes spin, and ``L" denotes LapH eigenvector dilution.  The schemes
are denoted by 1 for no dilution, F for full dilution, and B$J$ and I$J$ for
block-$J$ and interlace-$J$, respectively.  For example, full time and spin
dilution with interlace-8 LapH eigenvector dilution is denoted by
(TF, SF, LI8).

\begin{figure}[t]
\includegraphics[width=3.2in,bb=17 32 573 454]{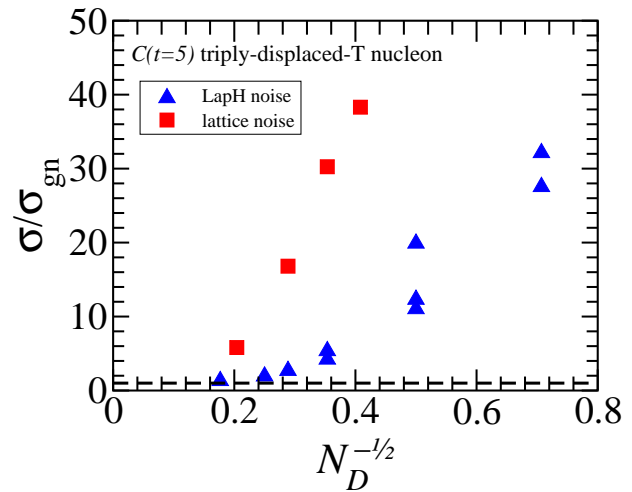}
\caption{Comparison of the new stochastic LapH method (triangles)
with the earlier stochastic method using noise on the
full lattice (squares) for the correlator $C(t=5)$ of a
triply-displaced-T nucleon operator on a $16^3\times 128$ lattice.
The vertical scale is the ratio of statistical error $\sigma$ (with no 
averaging over the six permutations of the three noises) over the
error in the gauge-noise limit $\sigma_{\rm gn}$, and in the horizontal
scale, $N_D$ is the number of Dirac-matrix inversions per source per
quark line.  Each point shows an error ratio using a particular dilution
scheme.  The LapH points lie significantly below the results from 
the lattice noise method, indicating a dramatic variance reduction.
\label{fig:lapholdnew}}
\end{figure}

Introducing noise in this
way produces correlation functions with significantly reduced variances,
as shown in Fig.~\ref{fig:lapholdnew}.  Let $C(t)$ denote the correlation function
of a representative triply-displaced-T nucleon operator at temporal separation 
$t$.  Let $\sigma_{\rm gn}$ represent the statistical error in $C(t=5)$ using
exactly-determined slice-to-slice quark propagators.  In other words, 
$\sigma_{\rm gn}$ arises solely from the statistical fluctuations in the gauge
configurations themselves (the gauge noise limit).  Let $\sigma$ denote the error
in $C(t=5)$ using stochastic estimates of the quark propagators.  The vertical 
axis in Fig.~\ref{fig:lapholdnew} is the ratio of the statistical error 
$\sigma$ in $C(t=5)$ over $\sigma_{\rm gn}$.  Results are shown for a variety of 
different dilution schemes.  In the lattice noise method, variance reduction
is achieved with projectors which dilute in the time, spin, and color indices.
Simple spatial dilutions are also used.  The squares show results for dilution
schemes with noise introduced in the larger spin-color-space-time vector
space, and the triangles show results for different dilution schemes
using noise introduced only in the LapH subspace.  The triangles show nearly an
order of magnitude reduction in the statistical error, compared to the square
symbols.  The correlator
for other time separations $t$ and for a variety of other hadron operators
were also examined.  All of the observables we studied showed the same 
dramatic reduction in the variance using the new LapH-noise method compared
to the lattice-noise method.

\begin{figure}[t]
\includegraphics[width=3.2in,bb=17 32 574 454]{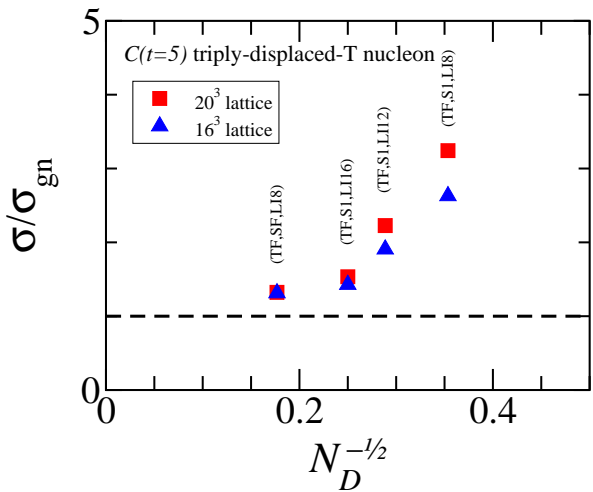}
\caption{Comparison of the new stochastic LapH method on $16^3$ 
(triangles) and $20^3$ (squares) lattices for the correlator $C(t=5)$ of a
triply-displaced-T nucleon operator on a $16^3\times 128$ lattice.
The vertical scale is the ratio of statistical error $\sigma$ (with 
averaging over the six permutations of the three noises) over the
error in the gauge-noise limit $\sigma_{\rm gn}$, and in the horizontal
scale, $N_D$ is the number of Dirac-matrix inversions per source per
quark line.  Each point shows an error ratio using a particular dilution
scheme.  The number of Laplacian eigenvectors needed is 32 on the 
$16^3$ lattice and 64 on the $20^3$ lattice.  The leftmost points
correspond to the dilution scheme (TF, SF, LI8).  For this scheme,
$\sigma/\sigma_{\rm gn}=1.31$ on the $16^3$ lattice and
$\sigma/\sigma_{\rm gn}=1.32$ on the $20^3$ lattice.
\label{fig:laphvol}}
\end{figure}

The number of Dirac matrix inversions needed in the stochastic LapH method
to achieve a target statistical precision was found to be 
insensitive to the spatial volume, despite the rapid increase in the number
of LapH eigenvectors.
Calculations on a $16^3$ and a $20^3$ lattice were carried out and the ratios
$\sigma/\sigma_{\rm gn}$ for various correlators at various time separations 
were compared.  The error ratios for the representative triply-displaced-T 
nucleon correlator for time separation $t=5$ on a $16^3$ lattice (triangles) 
are compared to those from a $20^3$ lattice (squares) in Fig.~\ref{fig:laphvol}. 
For the (TF, SF, LI8) dilution scheme, we found 
$\sigma/\sigma_{\rm gn}=1.31$ for this quantity on the $16^3$ lattice and
$\sigma/\sigma_{\rm gn}=1.32$ on the $20^3$ lattice.  Not only is the
equality of these ratios on the two volumes remarkable, but also their closeness
to unity is striking.  Keep in mind that the number of Laplacian eigenvectors
needed doubles in going from the smaller to the larger volume.  These results 
show that once a sufficient number of dilution projectors are used, the number
of inversions required by the stochastic LapH method does not increase with the
lattice volume and are sufficient to essentially reach the gauge noise limit.
Additional inversions of the Dirac matrix are totally unnecessary since they 
do not lower the error any further.  Other time separations and a variety of
other hadron operators were also studied and led to the same conclusions.

Different dilution schemes were explored using $16^3$, $20^3$, and $24^3$
spatial lattices with spacing $a_s\sim 0.12$~fm and light quark masses 
yielding pion masses ranging from 240~MeV to 500~MeV, and we have found that
the scheme (TF, SF, LI8) produces variances near that of the gauge noise limit
for correlators which involve only quark lines that connect the source and
sink time slices.  Interlace-$J$ and block-$J$ were observed to work equally 
well for spin and LapH eigenvector dilutions.  For correlators which involve 
quark lines that originate
and terminate at the final sink time, the dilution scheme (TI16, SF, LI8)
was found to work well.  The interlacing in time enables us to evaluate
quark lines that originate on \textit{any} time slice.  Results for 
several isoscalar correlators using (TI32, SF, LI8) on 20 configurations 
were compared with (TI16, SF, LI8) and no differences in the variances
were discernible, suggesting the gauge noise limit has essentially
been reached.

In the stochastic LapH method, the number of times that $\Omega x=y$ must be 
solved is $N_\rho N_P$ for each gauge-field configuration, where $N_\rho$
is the number of $Z_N$ noises used and $N_P$ is the number of dilution 
projectors. Using full time dilution (with four choices of source time $t_0$), 
full spin dilution and interlace-8 LapH eigenvector dilution, then the 
$t_0$-to-$t_F$ (for sink time $t_F$) quark lines require 128 inversions for 
each noise on each configuration.  To accommodate a baryon-meson system,
at least 5 noises for these quark lines are needed. The $t_F$-to-$t_F$ quark 
lines use interlace-16 time dilution, full spin dilution, and interlace-8 LapH
eigenvector dilution, requiring 512 inversions per noise per configuration.
Two noises are required for these quark lines.  Hence, a total of
$5\times 128+2\times 512=1664$ inversions per configuration are needed
to compute all baryons and mesons composed of $u,d$ quarks.  This number
of inversions is the same for both the $24^3$ and $32^3$ lattices that we
plan to use.  An exact treatment of the quark propagation requires
57,344 inversions per configuration on $24^3\times 128$ lattices for $N_v=112$
and 270,336 inversions per configuration on $32^3\times 256$ lattices
for $N_v=264$.

\section{Temporal correlations of hadron operators}
\label{sec:corr}

Details on how the temporal correlations of hadron operators are
evaluated using the stochastic LapH method are presented in this
section.  We limit our attention to four cases: baryon to baryon, 
meson to meson, two-meson to meson, and two-meson to two-meson
systems.  Other source to sink cases are straightforward
generalizations of the four examples below.

\subsection{Baryon to baryon correlations}

All of our hadrons are assemblages of basic building blocks
which are covariantly-displaced, LapH-smeared quark fields:
\begin{equation}
 q^A_{a\alpha j}= D^{(j)}\widetilde{\psi}_{a\alpha}^{(A)},
 \qquad  \overline{q}^A_{a\alpha j} = \widetilde{\overline{\psi}}_{a\alpha}^{(A)}
  \gamma_4\, D^{(j)\dagger},
\label{eq:quarkdef}
\end{equation}
where $a$ is a color index, $\alpha$ is a Dirac spin component,
$j$ is a displacement type, and $A$ is a quark flavor. 
To simplify notation, the Dirac spin component and the displacement
type are combined into a single index in what follows.

Each baryon operator destroying a three-momentum $\pvec$ is a 
linear superposition of so-called elemental three-quark operators, 
which are gauge-invariant terms of the form
\[
  \Phi^{ABC}_{\alpha\beta\gamma}(\pvec,t)= 
\sum_{\bm{x}} e^{-i\pvec\cdot\xvec}\varepsilon_{abc}
\, q^A_{a\alpha}(\bm{x},t)
\, q^B_{b\beta}(\bm{x},t)
\, q^C_{c\gamma}(\bm{x},t).
\]
The ``barred'' three-quark elemental operators which create a 
momentum $\pvec$ have the form
\[
  \overline{\Phi}_{\alpha\beta\gamma}^{ABC}(\pvec,t)= 
 \sum_{\bm{x}} e^{i\pvec\cdot\xvec}\varepsilon_{abc}
\ \overline{q}^C_{c\gamma}(\bm{x},t)
\ \overline{q}^B_{b\beta}(\bm{x},t)
\ \overline{q}^A_{a\alpha}(\bm{x},t). 
\]
We use hadron operators which transform irreducibly under all 
symmetries of the three-dimensional cubic lattice.
Each baryon sink operator, being a linear superposition of the three-quark
elemental operators, has the form
\begin{equation}
  B_{l}(t)= c^{(l)}_{
 \alpha\beta\gamma}\ \Phi^{ABC}_{\alpha\beta\gamma}(t),
\end{equation}
where $l$ is a compound index comprised of a three-momentum $\pvec$, an
irreducible representation (irrep) $\Lambda$ of the lattice symmetry group, the
row $\lambda$ of the irrep, isospin and other flavor quantum numbers, and
an identifier labeling the different operators in each symmetry channel.  
The corresponding source operators are
\begin{equation}
  \overline{B}_{l}(t)= c^{(l)\ast}_{
 \alpha\beta\gamma}\ \overline{\Phi}^{ABC}_{\alpha\beta\gamma}(t).
\end{equation}

The baryon correlation matrix elements are given by
\begin{equation}
  C_{l\overline{l}}(t_F\!-\!t_0)
  = \frac{1}{N_t}\sum_{t_0}\langle \  B_{l}(t_F)
   \ \overline{B}_{\overline{l}}(t_0) \ \rangle , 
\end{equation}
where $\langle\dots\rangle$ denotes a vacuum expectation value, which
is given by the usual ratio of path integrals over the fermion and gauge 
fields Wick-rotated into imaginary time.  
To simplify notation, we replace the average over all source times 
by a single fixed time $t_0$, exploiting time-translation invariance,
and obtain
\[
  C_{l \overline{l}}(t_F\!-\!t_0)
  = c^{(l)}_{\alpha\beta\gamma}
  c^{(\overline{l})\ast}_{
 \overline{\alpha}\overline{\beta}\overline{\gamma}}
\langle \ \Phi^{ABC}_{\alpha\beta\gamma}(t_F)
\ \overline{\Phi}^{\overline{A}\overline{B}\overline{C}}_{\overline{\alpha}
  \overline{\beta}\overline{\gamma}}(t_0)\ \rangle.
\]
Expand the three-quark elemental operators in terms of the 
covariantly-displaced smeared quark fields,
\begin{eqnarray*}
  && C_{l \overline{l}}(t_F\!-\!t_0)
   =  c^{(l)}_{\alpha\beta\gamma}
  c^{(\overline{l})\ast}_{
 \overline{\alpha}\overline{\beta}\overline{\gamma}}
\sum_{\bm{x}\overline{\bm{x}}}\ \varepsilon_{abc}
\   \varepsilon_{\overline{a}
 \overline{b}\overline{c}} e^{-i\pvec\cdot(\xvec-\overline{\xvec})}
   \nonumber\\
&&\times  \langle \, q^A_{a\alpha}(\xvec,t_F) \, q^B_{b\beta}(\xvec,t_F)
\, q^C_{c\gamma}(\xvec,t_F)
\nonumber\\
&&\qquad \times  \ \overline{q}^{\overline{C}}_{
 \overline{c}\overline{\gamma}}(\overline{\xvec},t_0)
\ \overline{q}^{\overline{B}}_{\overline{b}\overline{\beta}}
 (\overline{\xvec},t_0)
  \  \overline{q}^{\overline{A}}_{\overline{a}\overline{\alpha}}
(\overline{\xvec},t_0)\ \rangle,
\end{eqnarray*}
where the three-momenta associated with $l$ and $\overline{l}$ are assumed
to be the same $\pvec$, then evaluate the path integrals over the Grassmann 
fields to obtain a sum over products of quark lines, defining $t=t_F-t_0$:
\begin{eqnarray*}
  &&C_{l\overline{l}}(t)
  = c^{(l)}_{\alpha\beta\gamma}
  c^{(\overline{l})\ast}_{
 \overline{\alpha}\overline{\beta}\overline{\gamma}}
 \sum_{\bm{x}\overline{\bm{x}}}\ \varepsilon_{abc}
\   \varepsilon_{\overline{a}\overline{b}\overline{c}} 
  e^{-i\pvec\cdot(\xvec-\overline{\xvec})}\  \nonumber\\
&\times& \Bigl\langle {\cal Q}^{(A\overline{A})}_{
 a\alpha ;\overline{a}\overline{\alpha}}
{\cal Q}^{(B\overline{B})}_{b\beta ;\overline{b}\overline{\beta}}
{\cal Q}^{(C\overline{C})}_{c\gamma ;\overline{c}\overline{\gamma}}
-{\cal Q}^{(A\overline{A})}_{a\alpha;\overline{a}\overline{\alpha}}
{\cal Q}^{(B\overline{C})}_{b\beta   ;\overline{c}\overline{\gamma}}
{\cal Q}^{(C\overline{B})}_{c\gamma ;\overline{b}\overline{\beta}}
\nonumber\\
&-&{\cal Q}^{(A\overline{B})}_{a\alpha;\overline{b}\overline{\beta}}
{\cal Q}^{(B\overline{A})}_{b\beta  ;\overline{a}\overline{\alpha}}
{\cal Q}^{(C\overline{C})}_{c\gamma ;\overline{c}\overline{\gamma}}
- {\cal Q}^{(A\overline{C})}_{a\alpha;\overline{c}\overline{\gamma}}
{\cal Q}^{(B\overline{B})}_{b\beta ;\overline{b}\overline{\beta}}
{\cal Q}^{(C\overline{A})}_{c\gamma ;\overline{a}\overline{\alpha}}
 \nonumber\\
&+&
{\cal Q}^{(A\overline{C})}_{a\alpha;\overline{c}\overline{\gamma}}
{\cal Q}^{(B\overline{A})}_{b\beta ;\overline{a}\overline{\alpha}}
{\cal Q}^{(C\overline{B})}_{c\gamma ;\overline{b}\overline{\beta}}
+{\cal Q}^{(A\overline{B})}_{a\alpha;\overline{b}\overline{\beta}}
{\cal Q}^{(B\overline{C})}_{b\beta ;\overline{c}\overline{\gamma}}
{\cal Q}^{(C\overline{A})}_{c\gamma ;\overline{a}\overline{\alpha}}
\Bigr\rangle_U,
\end{eqnarray*}
where time and spatial labels have been omitted, and 
$\langle\dots\rangle_U$ is an expectation value defined as a ratio of path
integrals over the gauge field $U$ only, using the gauge field action
and the fermion determinant as the path integral weight.  Note that each
quark propagator ${\cal Q}$ connects each source site $\overline{\xvec}$
to each sink site $\xvec$, as well as connecting color and spin components
between the source and sink.  Hence, the summations in the above equation
are quite costly to carry out, and must be repeated over and over again
for every pair of baryon operators.

A dramatic simplification of the above equation can be achieved by using
Eq.~(\ref{eq:laphprop}) to estimate each quark line.
The following quantity emerges as a key component of the resulting
estimate:
\begin{eqnarray}
&&{\cal B}^{[b_1b_2b_3]}_{l}(\varphi_1,\varphi_2,\varphi_3;t)
 = c^{(l)}_{\alpha\beta\gamma}
 \sum_{\bm{x}}\ e^{-i\bm{p}\cdot\bm{x}}\varepsilon_{abc}\nonumber\\
&&\qquad\times \varphi^{[b_1]}_{a\alpha \bm{x}t}(\rho_1)
\ \varphi^{[b_2]}_{b\beta \bm{x}t}(\rho_2)
\ \varphi^{[b_3]}_{c\gamma \bm{x}t}(\rho_3),
\label{eq:baryonfunc}
\end{eqnarray}
where $b_1,b_2,b_3$ are noise dilution projector indices, and
the short-hand notation
$\varphi_k=\varphi(\rho_k)$ has been used, where $\varphi$ is the
quantity defined in Eq.~(\ref{eq:varphidef}).
The baryon correlation matrix element is then given by
\begin{eqnarray}
  C_{l \overline{l}}(t_F\!-\!t_0)&=&
\ \Bigl\langle
{\cal B}_{l}^{[b_1b_2b_3]}(\varphi_1,\varphi_2,\varphi_3;t_F)\nonumber\\
&\times&\Bigl( \delta_{ABC}^{\overline{A}\overline{B}\overline{C}}
\ {\cal B}_{\overline{l}
   }^{[b_1b_2b_3]}(\varrho_1,\varrho_2,\varrho_3;t_0)\nonumber \\
&-&\delta_{ABC}^{\overline{A}\overline{C}\overline{B}}
\ {\cal B}_{\overline{l}
   }^{[b_1b_3b_2]}(\varrho_1,\varrho_3,\varrho_2;t_0) \nonumber \\
&-&\delta_{ABC}^{\overline{B}\overline{A}\overline{C}}
\ {\cal B}_{\overline{l}
   }^{[b_2b_1b_3]}(\varrho_2,\varrho_1,\varrho_3;t_0) \nonumber \\
&-&\delta_{ABC}^{\overline{C}\overline{B}\overline{A}}
\ {\cal B}_{\overline{l}
   }^{[b_3b_2b_1]}(\varrho_3,\varrho_2,\varrho_1;t_0) \nonumber \\
&+&\delta_{ABC}^{\overline{C}\overline{A}\overline{B}}
\ {\cal B}_{\overline{l}
   }^{[b_2b_3b_1]}(\varrho_2,\varrho_3,\varrho_1;t_0) \nonumber \\
&+&\delta_{ABC}^{\overline{B}\overline{C}\overline{A}}
\ {\cal B}_{\overline{l}
   }^{[b_3b_1b_2]}(\varrho_3,\varrho_1,\varrho_2;t_0)
\Bigr)^\ast\Bigr\rangle_{U,\rho}
\label{eq:baryoncorr}
\end{eqnarray}
where $\delta_{ABC}^{DEF}=\delta_{AD}\delta_{BE}\delta_{CF}$ and
 $\langle\dots\rangle_{U,\rho}$ indicates an expectation value
over the gauge field $U$ and any $Z_N$ noises $\rho_k$.
Again, the above equation uses the short-hand notation
$\varphi_k=\varphi(\rho_k)$ and $\varrho_k=\varrho(\rho_k)$,
where the quark sinks $\varphi$ are defined in Eq.~(\ref{eq:varphidef})
and the quark sources $\varrho$ are defined in Eq.~(\ref{eq:varrhodef}).
$A,B,C$ are the quark flavors of the first, second, and third quarks
as ordered in the ${\cal B}$ functions.
To increase statistics, an average of the six permutations of the $1,2,3$ 
superscripts labeling the quark lines can be used, and if the masses of all 
three quark lines are the same, this requires no further inversions
of the Dirac matrix.

A very useful feature of Eq.~(\ref{eq:baryoncorr}) is the fact that the baryon
correlator completely factorizes into a function associated with the sink 
time slice $t_F$, and another function associated with the source time 
slice $t_0$.  Summations over color, spin, and spatial sites at the
source have been completely separated from the color, spin, and spatial
summations at the sink.  The stochastic LapH method leads to complete 
factorization of
hadron sources and sinks in temporal correlations, which greatly simplifies 
the logistics of evaluating correlation matrices involving large numbers
of operators.  
Eq.~(\ref{eq:baryoncorr}) also shows that implementing the Wick
contractions of the quark lines is also straightforward.  Contributions 
from different Wick orderings within a class of quark-line diagrams differ
only by permutations of the noises at either the source or the sink.
In Eq.~(\ref{eq:baryoncorr}), permutations of the noises at the source
have been used since this is generally much less costly.

\begin{figure}[t]
\includegraphics[width=3.2in,bb=0 20 490 200]{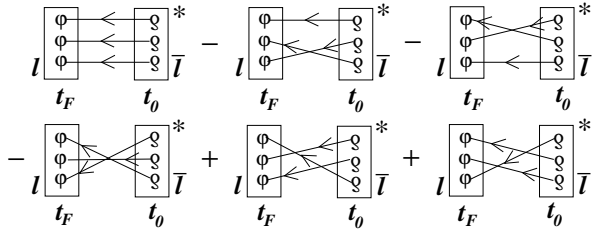}
\caption{Graphical depiction of Eq.~(\protect\ref{eq:baryoncorr}) for a 
baryon correlator $C_{l\overline{l}}(t_F\!-\!t_0)$ with source time $t_0$ 
and later sink time $t_F$.  Each box represents a baryon function given by
Eq.~(\protect\ref{eq:baryonfunc}) with the first quark located at the 
top of the box.  Lines connecting a $\varrho$ with a $\varphi$ indicate
summation over their dilution projector identifiers. The same noise must 
be used at the two ends of any single line, and different noises should be 
used for different lines.  Any line connecting quarks of different flavors 
represents a zero value.  The asterisks indicate complex conjugation.
\label{fig:baryoncorr}}
\end{figure}

Given the plethora of indices, a graphical representation of the above
equation is useful and is shown in Fig.~\ref{fig:baryoncorr}.  The quark
field $\psi$ is represented by a quark sink $\varphi$ or a $\overline{\varrho}$,
and $\overline{\psi}$ becomes a $\varrho$ or a $\overline{\varphi}$.  We 
represent a baryon given by Eq.~(\ref{eq:baryonfunc}) by a box containing the 
quark sources or sinks vertically aligned with the first quark on the left in
Eq.~(\ref{eq:baryonfunc}) located at the top of the box.  We use lines
connecting a $\varrho$ with a $\varphi$ (or a $\overline{\varrho}$ with
a $\overline{\varphi}$) to denote summation over the dilution indices
associated with the connected $\varrho$ and $\varphi$.  The same noise must
be used at the two ends of any single line, and different noises should be
used for different lines.

\subsection{Meson to meson correlations}

Each meson operator destroying a three-momentum $\pvec$ is a linear 
superposition of quark-antiquark elemental operators which are linear 
superpositions of gauge-invariant terms of the form
\begin{equation}
 \Phi^{AB}_{\alpha\beta}(t)= 
\sum_{\bm{x}}e^{-i\pvec\cdot(\xvec+\frac{1}{2}(\bm{d}_\alpha+\bm{d}_\beta))}
  \delta_{ab}  \ \overline{q}^A_{a\alpha}(\bm{x},t)\ q^B_{b\beta}(\bm{x},t),
\end{equation}
where $q,\overline{q}$ are defined in Eq.~(\ref{eq:quarkdef}),
$\bm{d}_\alpha, \bm{d}_\beta$ are the spatial displacements of the
$\overline{q},q$ fields, respectively, from $\xvec$,
$A,B$ indicate flavor, and $\alpha,\beta$ are compound indices
incorporating both spin and quark-displacement types.  The phase factor
involving the quark-antiquark displacements is needed to ensure proper
transformation properties under $G$-parity for arbitrary displacement types.
The ``barred'' operators  which create a momentum $\pvec$ then take the form
\begin{equation}
 \overline{\Phi}_{\alpha\beta}^{AB}(t)=
 \sum_{\bm{x}} e^{i\pvec\cdot(\xvec+\frac{1}{2}(\bm{d}_\alpha+\bm{d}_\beta))}
   \delta_{ab}\ \overline{q}^B_{b\beta}(\bm{x},t)\ q^A_{a\alpha}(\bm{x},t).
\end{equation}

\begin{figure}[t]
\includegraphics[width=3.2in,bb=0 20 400 107]{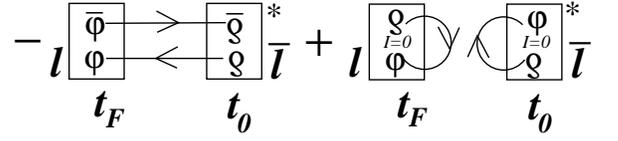}
\caption{Graphical depiction of Eq.~(\protect\ref{eq:mesoncorr}) for a 
meson correlator $C_{l\overline{l}}(t_F\!-\!t_0)$ with source time $t_0$ 
and later sink time $t_F$.  Each box represents a meson function given by
Eq.~(\protect\ref{eq:mesonfunc}) with the first quark field located at the 
top of the box.  Lines connecting a $\varrho$ with a $\varphi$ 
or a $\overline{\varrho}$ with a $\overline{\varphi}$ indicate
summation over their dilution projector identifiers. The same noise must 
be used at the two ends of any single line, and different noises should be 
used for different lines.  Any line connecting quarks of different flavors 
represents a zero value.  The asterisks indicate complex conjugation.
Contributions from the meson internal lines occur only for isoscalar mesons.
\label{fig:mesoncorr}}
\end{figure}

Each meson sink operator has the form
\begin{equation}
  M_{l}(t)= c^{(l)}_{\alpha\beta}\ \Phi^{AB}_{\alpha\beta}(t),
\end{equation}
(or is a flavor combination of the above form),
where again, the $l$ label includes the momentum $\pvec$, the symmetry
group irrep $\Lambda$, the row $\lambda$ of the irrep, and an identifier
specifying the different operators in each symmetry channel.  The 
corresponding source operators are
\begin{equation}
  \overline{M}_{l}(t)= c^{(l)\ast}_{
 \alpha\beta}\ \overline{\Phi}^{AB}_{\alpha\beta}(t).
\end{equation}

The meson correlation matrix elements are given by
\begin{equation}
  C_{l\overline{l}}(t_F\!-\!t_0)
  = \frac{1}{N_t}\sum_{t_0}\langle \  M_{l}(t_F)
   \ \overline{M}_{\overline{l}}(t_0) \ \rangle.
\end{equation}
In terms of the elemental operators and using translation invariance,
 the above equation becomes,
\begin{equation}
  C_{l \overline{l}}(t_F\!-\!t_0)= 
c^{(l)}_{\alpha\beta}
  c^{(\overline{l})\ast}_{
 \overline{\alpha}\overline{\beta}}
\langle \ \Phi^{AB}_{\alpha\beta}(t_F)
\ \overline{\Phi}^{\overline{A}\overline{B}}_{\overline{\alpha}
  \overline{\beta}}(t_0) \ \rangle,
\end{equation}
using translation invariance to fix to a single $t_0$ for notational
convenience.  Expand the elemental operators in terms of the 
covariantly-displaced smeared quark fields:
\begin{eqnarray*}
 && C_{l \overline{l}}(t_F\!-\!t_0) =  c^{(l)}_{\alpha\beta}  
  c^{(\overline{l})\ast}_{
 \overline{\alpha}\overline{\beta}}\sum_{\bm{x}\overline{\bm{x}}}
 e^{-i\pvec\cdot(\xvec+\frac{1}{2}(\bm{d}_\alpha+\bm{d}_\beta))}\\
 &&\quad\times e^{i\pvec\cdot(\overline{\xvec}+\frac{1}{2}(
  \bm{d}_{\overline{\alpha}}
      +\bm{d}_{\overline{\beta}}))} 
  \langle \  \overline{q}^A_{a\alpha}(\xvec,t_F)
 \  q^B_{a\beta}(\xvec,t_F)\\
&&\qquad \times
\ \overline{q}^{\overline{B}}_{\overline{a}\overline{\beta}}(
\overline{\xvec},t_0)
\ q^{\overline{A}}_{\overline{a}\overline{\alpha}}(\overline{\xvec},t_0) 
\ \rangle,
\end{eqnarray*}
where the three-momenta associated with $l$ and $\overline{l}$ are assumed
to be the same $\pvec$.
Next, the path integrals over the Grassmann fields are carried
out, and one obtains, for $t=t_F-t_0$,
\begin{eqnarray}
  &&C_{l \overline{l}}(t)
  = c^{(l)}_{\alpha\beta} c^{(\overline{l})\ast}_{
 \overline{\alpha}\overline{\beta}}\sum_{\bm{x}\overline{\bm{x}}}
 e^{-i\pvec\cdot(\xvec+\frac{1}{2}(\bm{d}_\alpha+\bm{d}_\beta))}\nonumber\\
 &&\quad\times e^{i\pvec\cdot(\overline{\xvec}+\frac{1}{2}(
  \bm{d}_{\overline{\alpha}}
      +\bm{d}_{\overline{\beta}}))} 
 \Bigl\langle -{\cal Q}^{(\overline{A}A)}_{\overline{a}
\overline{\alpha}; a\alpha }
\, {\cal Q}^{(B\overline{B})}_{a\beta ;\,\overline{a}
\overline{\beta}}\nonumber\\
 &&\quad+{\cal Q}^{(BA)}_{a\beta; a\alpha }
\, {\cal Q}^{(\overline{A}\overline{B})}_{\overline{a}\overline{\alpha}
  ;\,\overline{a}\overline{\beta}}
 \Bigr\rangle_U,
\end{eqnarray}
omitting time and spatial labels.  Eq.~(\ref{eq:laphprop}) or
Eq.~(\ref{eq:laphpropg5}) can then be used to estimate the two 
quark propagators.
In the first term, we find that it is advantageous to use 
Eq.~(\ref{eq:laphpropg5}) for the $A$ quark line and Eq.~(\ref{eq:laphprop}) 
for the $B$ quark line.

\begin{figure}[t]
\includegraphics[width=3.2in,bb=0 20 398 167]{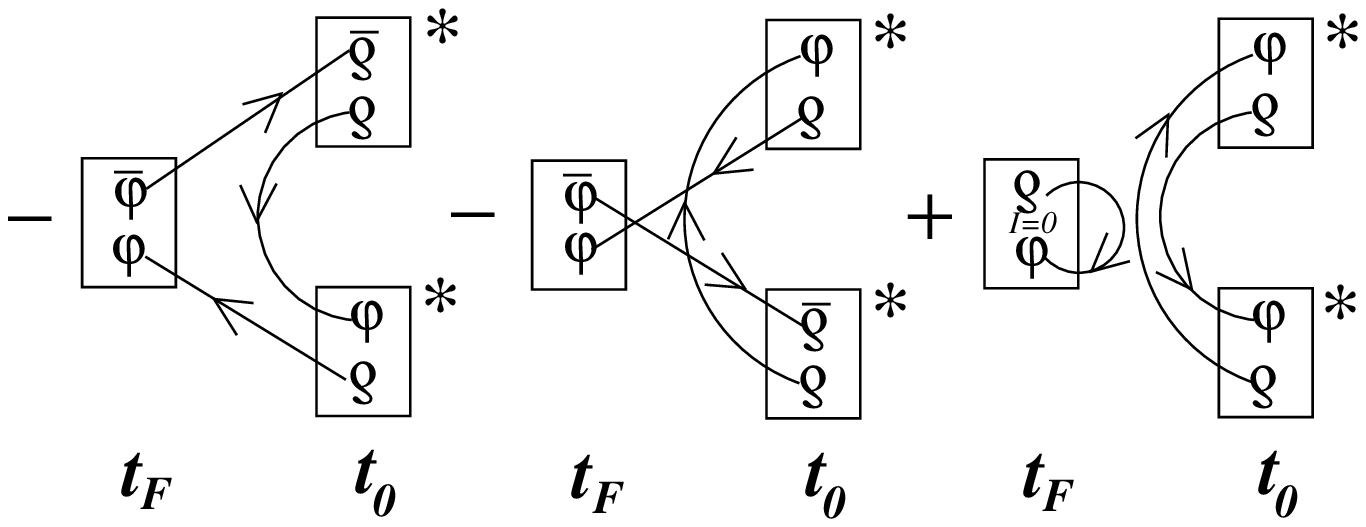}
\caption{Computation of the temporal correlation of a two-meson source
at time $t_0$ and a single-meson sink at time $t_F$.  The source mesons
are assumed to be non-isoscalars.  Each box represents a meson function given 
by Eq.~(\protect\ref{eq:mesonfunc}) with the first quark field located at the 
top of the box.  Lines connecting a $\varrho$ with a $\varphi$ 
or a $\overline{\varrho}$ with a $\overline{\varphi}$ indicate
summation over their dilution projector identifiers. The same noise must 
be used at the two ends of any single line, and different noises should be 
used for different lines.  Any line connecting quarks of different flavors 
represents a zero value.  The asterisks indicate complex conjugation.
The diagram with an internal line contributes only for isoscalar mesons.
\label{fig:twomesonmeson}}
\end{figure}

To proceed, define the following meson function:
\begin{eqnarray}
&&{\cal M}^{[b_1b_2]}_{l}(\varrho_1,\varphi_2;t)
 = c^{(l)}_{\alpha\beta}
 \sum_{\bm{x}}\ e^{-i\pvec\cdot(\xvec+\frac{1}{2}(\bm{d}_\alpha+\bm{d}_\beta))}
 \nonumber\\
&&\qquad\qquad\times \varrho^{[b_1]}_{a\alpha \bm{x}t}(\rho_1)^\ast
\ \varphi^{[b_2]}_{a\beta \bm{x}t}(\rho_2),
\label{eq:mesonfunc}
\end{eqnarray}
where $b_1,b_2$ are noise dilution projector indices, and
the short-hand notation $\varphi_k=\varphi(\rho_k)$ has been used again.
The meson correlator is given by
\begin{eqnarray}
  &&C_{l \overline{l}}(t_F\!-\!t_0)\nonumber\\
&=&\ \Bigl\langle -\delta_{AB}^{\overline{A}\overline{B}}
\ {\cal M}_{l}^{[b_1b_2]}(\overline{\varphi}_1,\varphi_2;t_F)
\ {\cal M}_{\overline{l}
   }^{[b_1b_2]}(\overline{\varrho}_1,\varrho_2;t_0)^\ast\nonumber\\
&+& \delta_{A\overline{A}}^{B\overline{B}}
\ {\cal M}_{l}^{[b_1b_1]}(\varrho_1,\varphi_1;t_F)
\ {\cal M}_{\overline{l}
   }^{[b_2b_2]}(\varphi_2,\varrho_2;t_0)^\ast\Bigr\rangle_{U,\rho}
\label{eq:mesoncorr}
\end{eqnarray}
where $\delta_{AB}^{CD}=\delta_{AC}\delta_{BD}$.  The second term only
contributes to isoscalar mesons.  Again, color, spin, and spatial summations
at the sink have completely factorized from the summations at the source.
This equation is graphically represented 
in Fig.~\ref{fig:mesoncorr}.

\begin{figure}[t]
\includegraphics[width=3.2in,bb=0 20 439 496]{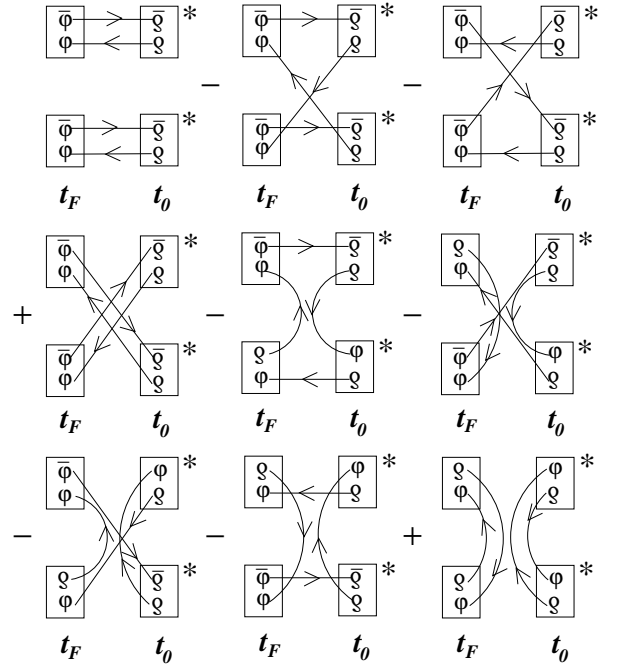}
\caption{Computation of the temporal correlation of a two-meson
source at time $t_0$ with a two-meson sink at time $t_F$.  All four
mesons are assumed to be non-isoscalars. Each box 
represents a meson function given by
Eq.~(\protect\ref{eq:mesonfunc}) with the first quark field located at the 
top of the box.  Lines connecting a $\varrho$ with a $\varphi$ 
or a $\overline{\varrho}$ with a $\overline{\varphi}$ indicate
summation over their dilution projector identifiers. The same noise must 
be used at the two ends of any single line, and different noises should be 
used for different lines.  Any line connecting quarks of different flavors 
represents a zero value.  The asterisks indicate complex conjugation.
\label{fig:twomesontwomeson}}
\end{figure}

\subsection{More complicated correlations}

The graphical rules developed in the preceding sections can be applied
to more complicated correlation matrix elements.  The correlation
of a two meson source with a single-meson sink is represented 
in Fig.~\ref{fig:twomesonmeson}.  The source mesons are assumed to be
non-isoscalars, otherwise there would be additional diagrams involving
meson internal lines.  The correlation of a two-meson source with a 
two-meson sink is illustrated in Fig.~\ref{fig:twomesontwomeson}.  All
four mesons are assumed to be non-isoscalars.  We apply $\gamma_5$ 
Hermiticity only in cases where a $\psi(t_0)$ at the source connects with 
a $\overline{\psi}(t_F)$ at the sink.  Full time dilution is the best 
choice for all quark lines connecting $t_0$ and $t_F$ and $t_0$ to $t_0$.
For all $t_F$-to-$t_F$ quark lines, interlacing in source time must 
be used.

To evaluate any correlation matrix element using the stochastic LapH method,
one first must identify the various hadron functions that are needed and
calculate them using Eqs.~(\ref{eq:baryonfunc}) and (\ref{eq:mesonfunc}).
These can be evaluated for a large set of hadron operators
and stored on disk.  The quark propagators are needed only at this stage
of the computation.  All color contractions and spatial sums
are carried out in evaluating the hadron functions.  Each hadron function
for a given choice of noises takes up very little space on disk since each
is an array over time and dilution indices only.  The correlation matrix
elements are then combinations of the different hadron functions for
different noise selections.  These final contractions involve only 
summations of dilutions indices.  In this way, a large number of correlation
matrices can be evaluated very efficiently.

\section{Implementation details}
\label{sec:implement}

Our software is written in C++ and links to the USQCD QDP++/Chroma
library\cite{Edwards:2004sx}.  Parts of our computations must be done using the
full four-dimensional lattice, but other parts are best handled
time slice by time slice in three dimensions.  QDP++ does not handle both 
three and four dimensional lattices simultaneously, so the different
parts of the computations were done in separate runs using both 3d and 4d
versions of our software.  Special input/output routines were written to
enable 4d QDP++ to read and write 3d time slices of the lattice.

Our computations are done as a sequence of tasks for each gauge configuration
in the Monte Carlo ensemble.  In the first task, the spatial links of the
gauge configuration are smeared using the stout-link procedure.  This
task is done using a four-dimensional version of our software, but the
smeared spatial links are written to disk as individual time slices suitable for
input to the three-dimensional version of our software.  In the second task,
computation of the Laplacian eigenvectors is done time slice by time slice in
three dimensions.  In the third task, the eigenvectors for the different
time slices are reorganized into four-dimensional eigenvectors corresponding
to the different eigenvalues.  The fourth task is the computation of the
quark propagators.  The inversions of the Dirac matrix must be done using
the full four-dimensional lattice, but our results are written to disk
once again as three-dimensional time slices.  Formation of the hadron
sources and sinks is accomplished in the fifth task using the
three-dimensional version of our software.  All of our hadron operators have
definite three momenta which involve summations over all spatial sites of
the lattice, so the resulting hadron sources and sinks are no longer lattice 
quantities. The final task is the assembly of the hadron sinks and sources 
to form the hadron correlation functions which can be accomplished using
a serial version of our software.

The eigenvectors of the gauge-covariant Laplacian are evaluated using
a Krylov-Spectral Restarted Lanczos (KSRL) method which is a modification of
the thick restarted Lanczos method described in Ref.~\cite{WuSimon}.
Let $A$ denote a Hermitian matrix whose lowest-lying or highest-lying 
eigenvectors are sought.
Given a starting vector $u$, the KSRL method begins by constructing
a Krylov space spanned by vectors $u,Au,A^2u,\dots,A^mu$.  The submatrix
of $A$ defined in this basis is then diagonalized, and the eigenvalues
and eigenvectors of this submatrix, known as the Ritz values and Ritz
vectors, are approximations to those of
the full matrix $A$.  Convergence to the exact eigenpairs occurs
as the Krylov space dimension increases, but a better procedure is
to stop the growth of the Krylov space at some point, typically
just above the number of desired eigenpairs, and restart the
procedure using a different starting vector or vectors.  The use of
a certain number of Ritz vectors to restart the procedure is known
as Krylov-Spectral restarting.  Key issues in the method are determining
how many Ritz vectors to use in restarting, determining the size of the
Krylov space to use, and maintaining orthogonality of the Lanczos 
vectors in finite-precision mathematics.

\begin{figure*}
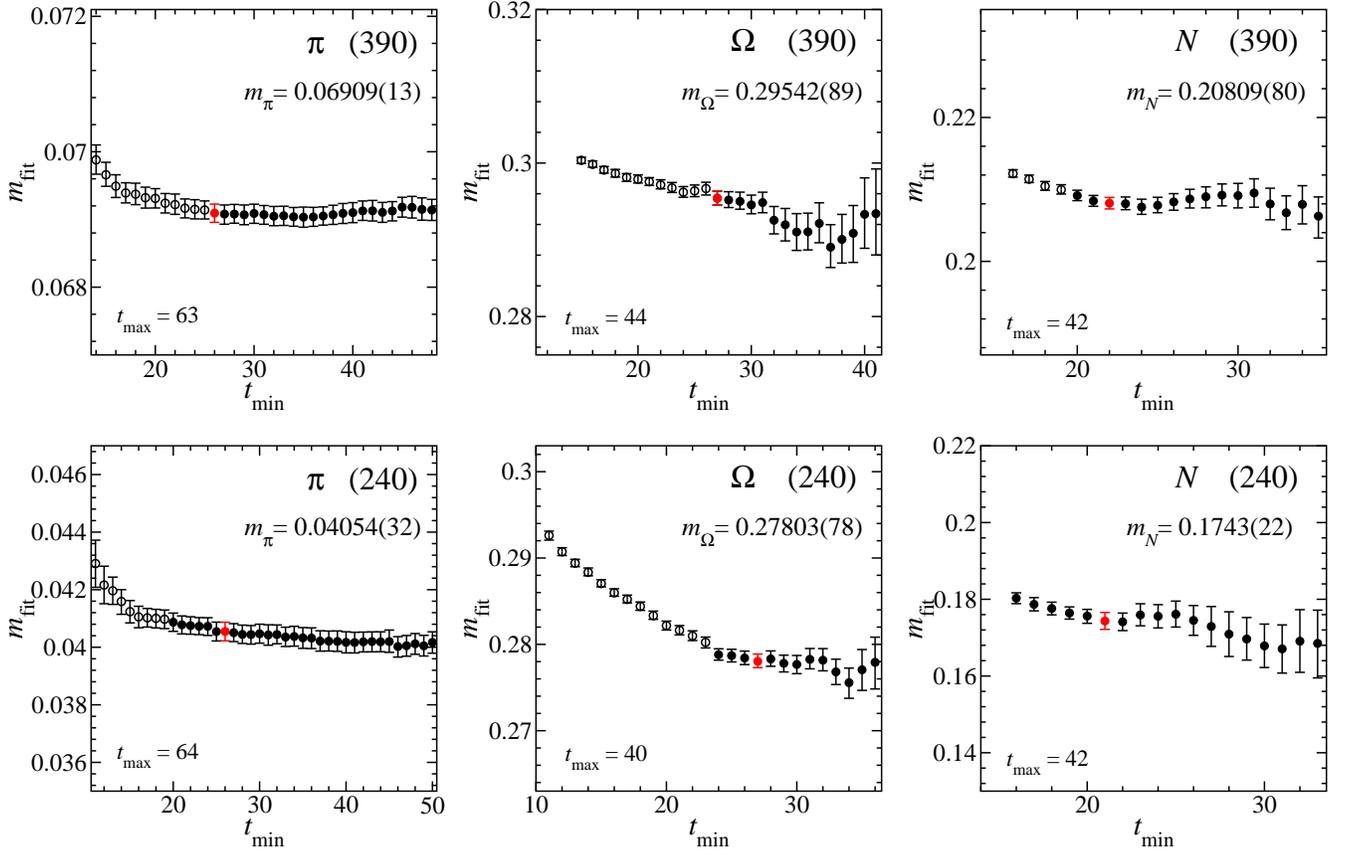

\begin{center}
\begin{minipage}{7.0in}
\includegraphics[width=2.2in,bb=20 22 541 532]{tmin840.pion.eps}\quad
\includegraphics[width=2.2in,bb=20 22 541 532]{tmin840.omegabaryon.eps}\quad
\includegraphics[width=2.2in,bb=20 22 541 532]{tmin840.nucleon.eps}\\[3mm]
\includegraphics[width=2.2in,bb=20 22 541 532]{tmin860.pion.eps}\quad
\includegraphics[width=2.2in,bb=20 22 541 532]{tmin860.omegabaryon.eps}\quad
\includegraphics[width=2.2in,bb=20 22 541 532]{tmin860.nucleon.eps}\\[-8mm]
\end{minipage}
\end{center}
\caption{
Masses $m_{\rm fit}(t)$ obtained by fitting the correlators for
single-site $\pi, \Omega, N$ operators to a cosh(exponential)
form for the $\pi(\Omega,N)$ in the temporal range 
$t_{\rm min}$ to $t_{\rm max}$.  Results
are shown for different $t_{\rm min}$ with $t_{\rm max}$ fixed to the
value stated in the lower left corner of each plot.  Open symbols
indicate unacceptable fit qualities, and solid symbols show results
with acceptable fit qualities $Q$.  The top row shows results using
551 configurations of the 390 ensemble on a $24^3\times 128$ lattice, 
and the bottom row shows results with 584 configurations of the 240 
ensemble on a $24^3\times 128$ lattice.  
The fit value given in each plot corresponds to the fit indicated by 
the red point.
\label{fig:scalesetting}}
\end{figure*}

In our calculations, we use either a random vector or the vector whose
components are all equal for the starting vector.  Full global 
reorthogonalization is used at all steps.  The decision to reorthogonalize 
multiple times is based on a simple criterion\cite{rutishauser}:  if the 
norm of the vector decreases by $1/\kappa$, where 
$\kappa = \sqrt{2}$, then further reorthgonalization is done.  A maximum of four
reorthogonalizations is enforced.  Equation~5 in Ref.~\cite{WuSimon2}
is used to choose the number of Ritz vectors to keep, except that
the number must be at least as large as the number of converged
vectors and cannot exceed the dimension of the Krylov space minus the
number of converged and locked vectors minus twelve.  For an approximate
eigenvector $x$ (with unit norm) and an estimate $\lambda$ of its corresponding 
eigenvalue, the residual norm is defined by 
$r=\vert\vert Ax-\lambda x\vert\vert$. An eigenpair is considered converged
when $r< \texttt{tol}\vert\vert A\vert\vert$, where $\texttt{tol}$ is the 
desired tolerance and the matrix 2-norm is defined by
$\vert\vert A\vert\vert = \max_{x\neq 0}\vert\vert Ax\vert\vert/\vert
\vert x\vert\vert$,
and can be estimated by the largest absolute value of any Ritz value encountered
in the computation.

In calculating the eigenvectors of $\widetilde{\Delta}$, Chebyshev 
acceleration is used. The eigenvalues of $-\widetilde{\Delta}$ are all real
and lie between 0 and some maximum value denoted by $\lambda_L$. We wish to
determine the eigenvectors corresponding to the lowest-lying eigenvalues 
lying between 0 and some cutoff $\lambda_C$.  The rate of convergence to 
solution increases with the spacing between the levels.  Convergence is much
faster for widely spaced levels.  Hence, convergence can be accelerated by 
transforming the spectrum so that the desired part of the spectrum is more
widely spaced.  The following transformation is applied first: 
\begin{equation}
     B = 1 + \frac{2}{(\lambda_L-\lambda_C)}\Bigl(\widetilde{\Delta}
  +\lambda_C\Bigr).
\end{equation}
The above transformation maps the unwanted spectrum to the range $-1\cdots 1$,
and the desired part lies above 1.  Chebyshev polynomials are then applied:
\begin{equation}
   A = T_n(B).
\end{equation}
Eigenvalues lying between -1 and 1 stay between -1..1, and the desired 
eigenvalues above 1 get spaced out to large and widely-separated
values above 1. The lowest-lying eigenvalue of 
$-\widetilde{\Delta}$ becomes the highest-lying eigenvalue of $A$.
Transforming the desired levels to the region above 1 is most convenient
since it allows the use of Chebyshev polynomials of any order, both even
and odd.  The Chebyshev polynomials are applied using the following 
recurrence relation: 
\begin{eqnarray}
   T_0(x) &=& 1,\qquad    T_1(x) = x,  \\
   T_n(x) &=& 2x\ T_{n-1}(x) - T_{n-2}(x).
\end{eqnarray}

For calculations done on our anisotropic $24^3\times 128$ lattices, we
need to compute the lowest-lying $N_v=112$ eigenvectors on each time slice.
A Krylov space dimension of 160 was found to work well, and $\lambda_L=15$ 
and $\lambda_C=0.5$ were appropriate.  Chebyshev polynomials of order $8$
were used, and the residual tolerance was set to $10^{-9}$.  Convergence
of all $N_v$ levels occurred within a dozen or less restarts.

The LapH eigenvectors are uniquely determined only to within an overall
phase.  Given the way in which $Z_N$ noise is injected in the LapH
subspace, one sees that a given quark line is not invariant
under a change of the phase multiplying each eigenvector (due to the
off-diagonal pieces not being exactly zero).  It turns out that changing
the phase is equivalent to changing the noise by a U(1) phase.
This is not a problem, but erroneous results can occur if the original 
eigenvector files used to determine the quark sinks get deleted and the
eigenvectors have to be reconstructed for making the hadrons.
With different run parameters, the eigensolver could produce a different 
phase.  The introduction of a phase convention eliminates this potential 
problem.

Once the needed eigenvectors of the Laplacian are computed and stored,
the next step is to compute all elements of 
$V_s^\dagger \Omega^{-1}V_s P^{[b]}\rho^{r}$.  There are only $N_tN_v$ 
elements to store for each noise $r$ and each dilution projector $b$, so
storage of these quark propagation coefficients is modest. Disk storage 
is actually dominated by the LapH eigenvectors.  Another nice
feature is the fact that the quark propagation coefficients are
gauge invariant, as long as the eigenvector phases are handled
appropriately.  Solving $\Omega x=y$ for $x$ with $y=V_sP^{[b]}\rho^{r}$ is 
accomplished
using a mixed-precision improved version of the biconjugate gradient method 
with even-odd preconditioning.  This was found to be the fastest inverter 
available in Chroma.  Occasionally convergence is not achieved, and a 
slower conjugate gradient solver is applied to the system 
$\Omega^\dagger \Omega x=\Omega^\dagger y$.  

\begin{figure}
\begin{center}
\includegraphics[width=3.2in,bb=23 47 523 531]{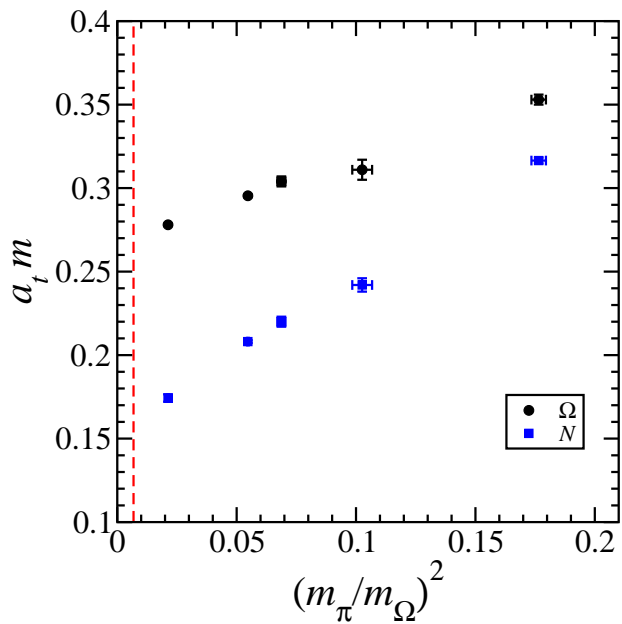}
\end{center}
\caption{
Products of $a_t$ and the nucleon and $\Omega$-baryon masses against 
$(m_\pi/m_\Omega)^2$ for fixed $\beta=1.5, m_s=-0.0743$ and varying 
$m_u=m_d$. The two leftmost points for each baryon are from this work,
and the three rightmost points are from Ref.~\protect\cite{Lin:2008pr}.
The vertical dashed line indicates the physical value of 
$(m_\pi/m_\Omega)^2$.
\label{fig:nucomega}}
\end{figure}

Our correlator estimates and their variances are insensitive to the value
of $N$ used for the $Z_N$ noise, as long as $N$ is not too small.  We found
that $N=4$ produced results indistinguishable in quality from those of larger
$N$. Hence, we use $Z_4$ noise in all of our computations.
We identify a $Z_4$ noise vector for an ensemble of gauge configurations
by a 16-bit unsigned integer $s$.  To create a noise vector $\rho^{(s)}$
for a gauge configuration labeled by an RHMC trajectory number $k$
(assumed to have a value ranging from 0 to $2^{16}-1$), a 32-bit unsigned 
integer $m$ is first formed in a particular manner using the 16 
binary digits of $s$ and the 16 bits of $k$.  Although the procedure of 
forming $m$ is arbitrary, the same procedure must be used in every instance. 
The 32-bit unsigned integer $m$ is then taken as a seed to the 32-bit 
Mersenne twister random number generator 
which is used to create the $Z_4$ noise $\rho^{(s)}(t,i,\alpha)$ for each
LapH eigenvector, labeled by time $t$ and level $i$, and for each spin 
index $\alpha$.  The elements of $\rho^{(s)}$ are generated in a 
particular order that is always the same.  Each $Z_4$ element is chosen
using the sequence of bits obtained from the current state of the Mersenne
twister, taking two bits at a time.  It was found that the linear 
congruential generator in QDP++/Chroma is not adequate for generating the
$Z_4$ noise and leads to serious errors in some instances.

\begin{figure*}
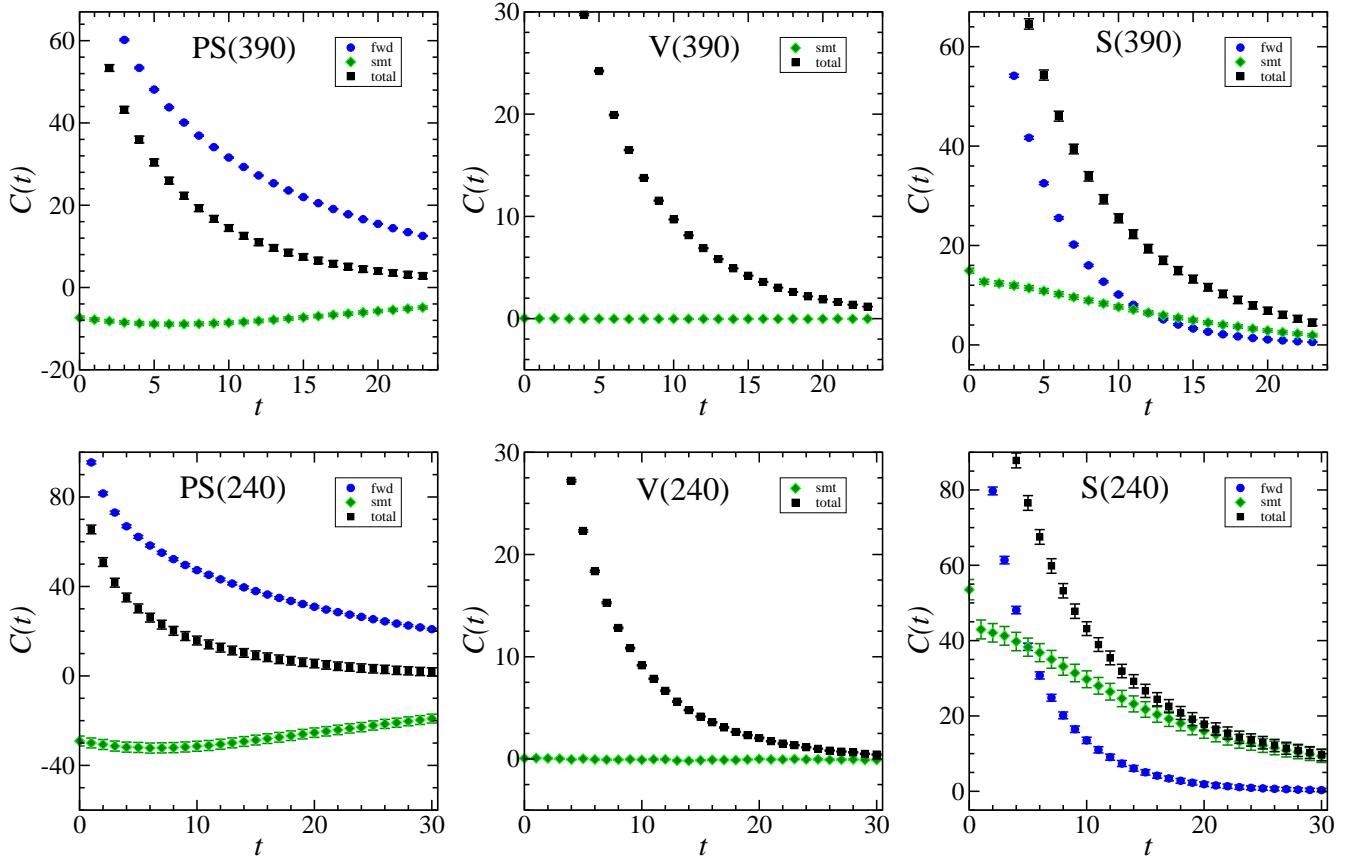

\begin{center}
\begin{minipage}{7.0in}
\includegraphics[width=2.2in,bb=21 34 523 530]{corr840.PS.eps}\quad
\includegraphics[width=2.2in,bb=21 34 523 530]{corr840.V.eps}\quad
\includegraphics[width=2.2in,bb=21 34 523 530]{corr840.S.eps}\\[3mm]
\includegraphics[width=2.2in,bb=21 34 523 530]{corr860.PS.eps}\quad
\includegraphics[width=2.2in,bb=21 34 523 530]{corr860.V.eps}\quad
\includegraphics[width=2.2in,bb=21 34 523 530]{corr860.S.eps}\\[-8mm]
\end{minipage}
\end{center}
\caption{
Correlators $C(t)$ against temporal separation $t$ for single-site 
operators which produce the isoscalar pseudoscalar (PS), vector (V), and 
scalar (S) mesons.  Results in the top row were obtained using 210
configurations (135 for the scalar channel) of the 390 ensemble.
Results in the bottom row were obtained using 198 configurations
of the 240 ensemble.  In the legends, ``fwd" refers to contributions
from the diagram containing only forward-time source-to-sink quark lines,
``smt" refers to contributions from the diagram containing only
quark lines that originate and terminate at the same time.  For the 
scalar channel, the ``smt" contribution has a vacuum expectation value
subtraction.  Forward-time quark lines use dilution scheme (TF, SF, LI8)
and same-time quark lines use (TI16, SF, LI8).  The lattice size is
$24^3\times 128$ for all the results shown here.
\label{fig:isoscalarscorr}}
\end{figure*}

\section{Initial applications}
\label{sec:testing}

Our initial development of the stochastic LapH method was done using 
a small $16^3$ spatial lattice which is not very interesting for
hadron physics.  Since the main reason for pursuing the method is to
apply it on large lattices for both single-hadron and multi-hadron
correlators, we proceeded to test the method by studying several
simple hadronic systems requiring sink-to-sink quark lines on a 
reasonably large $24^3\times 128$ anisotropic lattice having spatial 
volume (3~fm)$^3$. 

\begin{figure*}
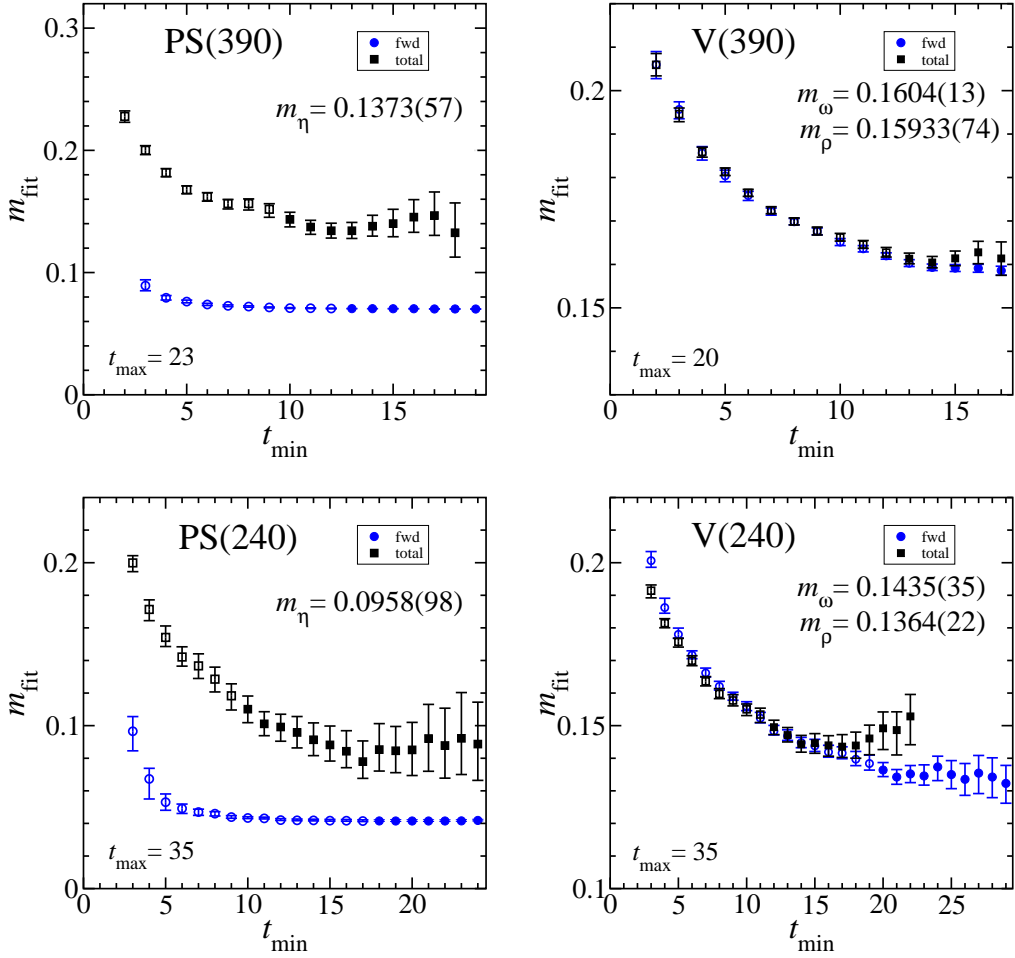

\begin{center}
\begin{minipage}{7.0in}
\includegraphics[width=2.5in,bb=21 34 543 530]{tminfits840.PS.eps}\qquad
\includegraphics[width=2.5in,bb=21 34 543 530]{tminfits840.V.eps}\\[5mm]
\includegraphics[width=2.5in,bb=21 34 543 530]{tminfits860.PS.eps}\qquad
\includegraphics[width=2.5in,bb=21 34 543 530]{tminfits860.V.eps}\\[-8mm]
\end{minipage}
\end{center}
\caption{
Masses $m_{\rm fit}(t)$ obtained by fitting the correlators $C(t)$ shown
in Fig.~\protect\ref{fig:isoscalarscorr} to a cosh
form in the temporal range $t_{\rm min}$ to $t_{\rm max}$.  Results
are shown for different $t_{\rm min}$ with $t_{\rm max}$ fixed to the
value stated in the lower left corner of each plot.  Open symbols
indicate unacceptable fit qualities, and solid symbols show results
with acceptable fit qualities $Q$.  The top row corresponds to the
390 ensemble, and the bottom row corresponds to the 240 ensemble.
The left-hand-side plots show results for the $\eta$ and $\pi$ pseudoscalar
mesons, and the right-hand-side plots show results for the $\omega$ and 
$\rho$ vector mesons.  The scalar channel is not shown here since a
reliable extraction of the lowest-lying energy in this channel needs
a two-pion operator. The lattice size is $24^3\times 128$ for all the 
results shown here.
\label{fig:isoscalarstmin}}
\end{figure*}

Two ensembles of gauge configurations were used.  These ensembles were
generated using the Rational Hybrid Monte Carlo (RHMC) 
algorithm\cite{Clark:2004cp}, which is a variant of the hybrid 
molecular-dynamics (HMC) algorithm\cite{Duane:1987de} suitable for
$N_f=2+1$ quark flavors.  The updating algorithm is a Metropolis method 
with a sophisticated means of proposing a global change to the gauge
and pseudofermion fields.  A fictitious momentum is introduced for each
link variable with a Gaussian distribution and Hamilton's equations involving
these momenta and the original action as a potential energy are 
approximately solved for some length of fictitious time, known as an
RHMC trajectory.  An improved anisotropic clover fermion action and
an improved gauge field action were used\cite{Lin:2008pr}.
In both ensembles, $\beta=1.5$ is used and the $s$ quark mass parameter 
is set to $m_s=-0.0743$
in order to reproduce a specific combination of hadron masses\cite{Lin:2008pr}.
In one ensemble, the light quark mass parameters are set to
$m_u=m_d=-0.0840$ so that the pion mass is around 390~MeV using one
particular way of setting the scale, discussed below. In the 
other ensemble, $m_u=m_d=-0.0860$ are used, resulting in a pion mass around
240~MeV.  We refer to these ensembles as the 390 and 240 ensembles, 
respectively.

We calculated the masses of the pion, the nucleon, and the $\Omega$-baryon.
Our results are shown in Fig.~\ref{fig:scalesetting} for the two ensembles 
on a $24^3\times 128$ lattice.  This figure demonstrates that the use
of our stochastic estimates of the smeared quark propagators still leads to
high accuracy results of standard quantities.  The nucleon and $\Omega$-baryon
masses times $a_t$ are shown in Fig.~\ref{fig:nucomega} against
$(m_\pi/m_\Omega)^2$.  Results from Ref.~\cite{Lin:2008pr} are also
included in this figure.  Our goal in this work is simply to present and
test the stochastic LapH method, so we defer a detailed analysis of these 
results until a later publication.  However, it is encouraging that 
fitting the three leftmost $\Omega$-baryon points to a form linear in
$(m_\pi/m_\Omega)^2$ and fitting the three leftmost nucleon points to
an empirical form linear in $m_\pi/m_\Omega$ yields 
$m_N/m_\Omega\approx 0.556$ at the physical value of $m_\pi/m_\Omega$,
which compares well with the observed 0.561 value. 

Our calculations determine all hadron masses in terms of the temporal
lattice spacing $a_t$.  In order to express the hadron masses in terms of
MeV, a value of $a_t^{-1}$ must be specified using an appropriate
renormalization scheme.  Away from the physical point, different
renormalization schemes will lead to different choices of $a_t^{-1}$.
One particular scheme that has been used in the past uses the
mass of the $\Omega$ baryon to set the scale when the strange quark
mass is close to the value that reproduces the physical value of
$(2m_K^2-m_\pi^2)/m_\Omega^2$.   Using this scheme, we find an inverse 
temporal spacing $a_t^{-1}= 5.661(17)$~GeV for the 390 ensemble and 
$a_t^{-1}=6.015(17)$~GeV for the 240 ensemble.  Since the ratio of 
spatial spacing over temporal spacing 
has been tuned to a value near 3.5, we have $a_s\sim 0.12$~fm for both of
these ensembles.  Our values for the pion and nucleon masses are 
$m_\pi=0.3911(14)$~GeV and $m_N=1.1781(58)$~GeV on the 390 ensemble, and
$m_\pi=0.2439(20)$~GeV and $m_N=1.048(14)$~GeV on the 240 ensemble.
An alternative scale-setting scheme would be to extrapolate the
$\Omega$-baryon mass results for different $m_u,m_d$ but fixed $\beta,m_s$
to the physical value of $m_\pi/m_\Omega$ using a form linear in 
$(m_\pi/m_\Omega)^2$ motivated by heavy baryon chiral perturbation theory,
then use the $\Omega$ mass to determine $a_t^{-1}$.  Doing this yields
$a_t^{-1}\sim 6.3$~GeV and pion masses 250 and 430~MeV for our two
ensembles.

\begin{figure*}
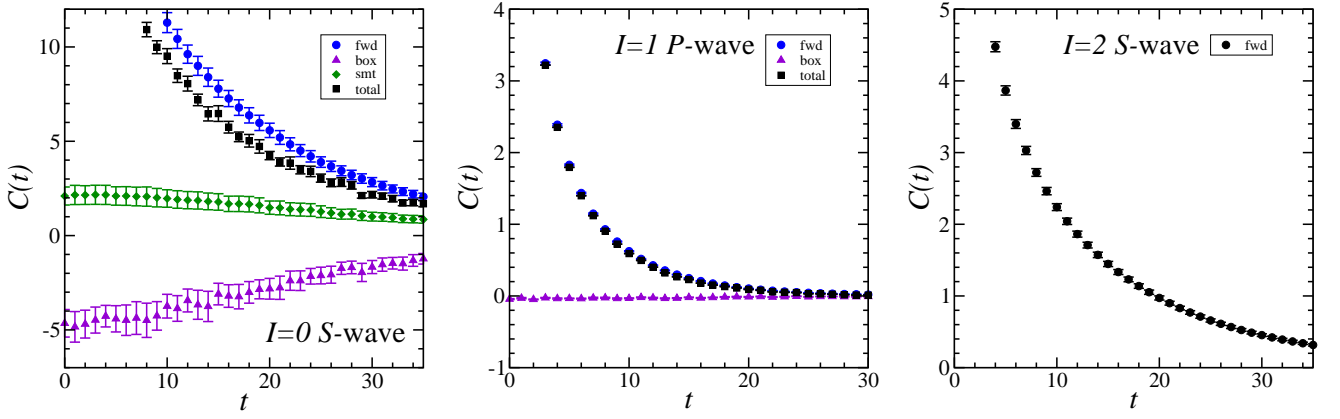

\begin{minipage}{7.0in}
\includegraphics[width=2.2in,bb=21 34 523 530]{pipi_pipi860_V_24_corr_I_0.eps}\quad
\includegraphics[width=2.2in,bb=21 34 523 530]{pipi_pipi860_V_24_corr_I_1.eps}\quad
\includegraphics[width=2.2in,bb=21 34 523 530]{pipi_pipi860_V_24_corr_I_2.eps}\\[-8mm]
\end{minipage}
\caption{
Correlators $C(t)$ against temporal separation $t$ for two-pion operators with
total isospin $I=0,1,2$ and zero total momentum.  $S$-wave results have zero 
relative momentum, $P$-wave has minimal non-zero on-axis relative momenta.
Results were obtained using 584 configurations of the 240 ensemble. In the legends,
 ``fwd" refers to contributions from diagrams containing only forward-time 
source-to-sink quark lines, ``smt" refers to contributions from diagrams 
containing only quark lines that originate and terminate at the same time, 
and ``box" refers to diagrams containing both kinds of quark lines.  
Forward-time quark lines use dilution scheme (TF, SF, LI8)
and same-time quark lines use (TI16, SF, LI8). The lattice size is 
$24^3\times 128$ for all the results shown here.
\label{fig:twopionscorr}}
\end{figure*}

\begin{figure}[b]
\includegraphics[width=3.2in,bb=10 22 540 522]{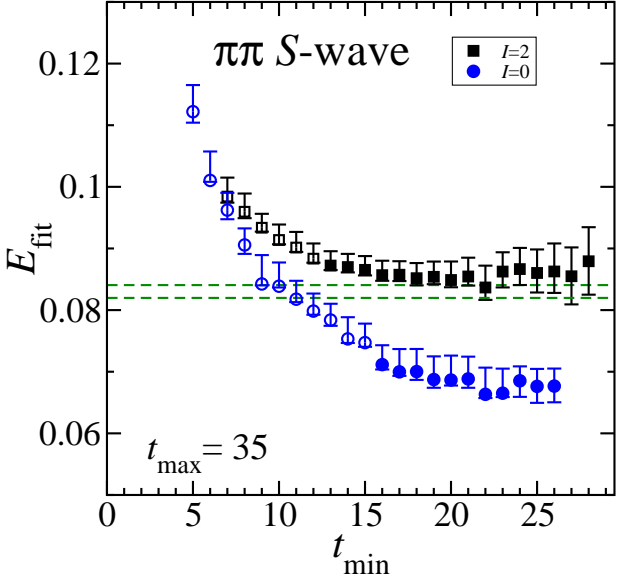}
\caption{
Energies $E_{\rm fit}(t)$ obtained by fitting the correlators $C(t)$ shown
in Fig.~\protect\ref{fig:twopionscorr} to a cosh + constant form in the 
temporal range $t_{\rm min}$ to $t_{\rm max}$.  Results
are shown for different $t_{\rm min}$ with $t_{\rm max}$ fixed to the
value stated in the lower left corner of the figure.  Open symbols
indicate unacceptable fit qualities, and solid symbols show results
with acceptable fit qualities $Q$.  These results were obtained using
584 configurations of the 240 ensemble.  The horizontal dashed lines
indicate the energy of two free pions at rest. The lattice size is 
$24^3\times 128$.
\label{fig:twopionstmin}}
\end{figure}

Results for the isoscalar mesons in the pseudoscalar, vector, and scalar
channels and the two-pion system of total isospin $I=0,1,2$ are presented in
Figs.~\ref{fig:isoscalarscorr}, \ref{fig:isoscalarstmin},
\ref{fig:twopionscorr}, and \ref{fig:twopionstmin}.  In these results, the 
dilution scheme (TF, SF, LI8) is used for all quark lines connecting source 
time $t_0$ to the sink time $t_F$ and $t_0$ to $t_0$.  Four widely-separated
source times $t_0$ were used on each gauge configuration.
For all $t_F$-to-$t_F$ quark lines, the dilution scheme (TI16, SF, LI8) is used.
Observables are evaluated using configurations
separated by $n_{\rm sep}$ RHMC trajectories, where $n_{\rm sep}=20$ for
the two-pion correlators and $n_{\rm sep}=40$ for the isoscalar meson 
correlators.  Jackknife binning shows autocorrelations to be suitably 
small.

Our goal here is simply to test the stochastic LapH method, so simple 
single-site operators involving only the light $u,d$ quarks are used for the
isoscalar mesons, and single-site pion operators are used to make the two-pion
states with zero and non-zero relative momenta.  The temporal correlations of 
such simple operators have significant contaminations from higher-lying states,
so that the effective masses associated with these correlations tend to a plateau
rather slowly.  Future work will make use of more sophisticated spatially-extended
operators.  The issue of mixing with $\overline{s}s$ operators is not addressed 
in these tests, and no vacuum-expectation-value subtraction is used for the 
$\eta$ correlator.  In chirally-symmetric fermion formulations, the expectation 
value of the unsmeared, isosinglet pseudoscalar operator would be proportional 
to the topological charge, which has notoriously long 
autocorrelation times and may not be sampled properly in a Monte Carlo 
simulation\cite{Brower:2003yx,Aoki:2007ka}.  This can show up 
as a non-zero vacuum expectation value for the $\eta$, which disappears
as the volume increases.  Our test results do not take such effects into 
account, but future work will investigate this.

In Fig.~\ref{fig:isoscalarscorr}, the contributions to the isoscalar temporal 
correlations $C(t)$ from the diagram containing only forward-time source-to-sink
quark lines are shown with label ``fwd", and the contributions from the diagram
containing only quark lines that originate and terminate at the same time are
shown with label ``smt".  The total correlator is also shown in each
case.  In the vector channel, the contribution from the same-time diagram
is very small and the total correlator can barely be distinguished from
the forward-line diagram contribution, so the ``fwd" contribution is not shown.
In the scalar channel, the accuracy of the ``smt" contribution is
particularly remarkable since a large vacuum expectation value has been
subtracted.

The correlators in Fig.~\ref{fig:isoscalarscorr} were used to extract
various isoscalar meson masses.  The pion and $\rho$ masses can be
obtained from the forward-line contributions to the pseudoscalar and
vector correlators, respectively.   
Correlated-$\chi^2$ fits to $A(e^{-mt}+e^{-m(N_t-t)})$ 
for temporal separations between $t_{\rm min}$ and $t_{\rm max}$ were done to
extract the masses of the particles.  Jackknife sampling was used to estimate 
the data covariance matrix, and bootstrap sampling was used to compute the
uncertainties in the fit parameters.  Results are shown in 
Fig.~\ref{fig:isoscalarstmin} for various $t_{\rm min}$ values, with 
$t_{\rm max}$ fixed to the value stated in each plot.
Even using such simple hadron operators, fairly accurate mass extractions 
are obtained.  Future use of better operators will certainly improve these
results.  Results are not shown for the scalar channel since the 
lowest-lying energy in this channel is a two-pion state.  Extractions of 
the energies in the scalar channel are best done with a correlator matrix
using both single-hadron and two-pion operators.  The excellent statistical
precision obtained for the correlators at small temporal separations 
suggests that diagonalizations of future correlation matrices 
estimated with stochastic LapH will be stable and accurate.

With current Monte Carlo algorithms on presently available computing
resources, it remains impractical to use light $u,d$ quark masses tuned
to properly reproduce the pion mass.  Hence, the $u,d$ quark masses
used here yield a pion mass which is too heavy, making comparison to
experiment somewhat problematical.  Using the $\Omega$-baryon mass to set
the inverse temporal spacing, we find masses $m_\eta=576(59)$~MeV, 
$m_\rho=820(13)$~MeV, and $m_\omega=863(21)$ using 198 configurations of
the 240 ensemble. The experimental values are $m_\eta=548$~MeV, 
$m_\rho=776$~MeV, and $m_\omega=783$~MeV.  Future work will use
better operators and all 584 configurations to achieve improved results.

Our results for the energies of two light pions are shown in
Figs.~\ref{fig:twopionscorr} and \ref{fig:twopionstmin}.  
We studied $S$-wave states of zero relative
momentum and total isospin $I=0$ and $I=2$, as well as a $P$-wave 
with minimal non-zero on-axis relative momenta in the $I=1$
channel. In Fig.~\ref{fig:twopionscorr}, contributions to the correlators 
from the diagrams containing only forward-time source-to-sink quark lines
are labeled by ``fwd", contributions from diagrams containing only quark
lines that originate and terminate at the same time are shown as ``smt",
and contributions labeled by ``box" are those from the diagrams containing
both kinds of quark lines (see Fig.~\ref{fig:twomesontwomeson}).  Energies
were extracted using correlated-$\chi^2$ fits to the form
$A+B(e^{-Et}+e^{-E(N_t-t)})$ in the range $t_{\rm min}$ to $t_{\rm max}$.
The results for different $t_{\rm min}$ are shown in Fig.~\ref{fig:twopionstmin},
for $t_{\rm max}$ fixed to the value stated in the figure.  Open symbols
indicate unacceptable fit qualities, whereas solid symbols indicate results
from fits of acceptable quality $Q$.  The constant term in the fit form 
arises from a source pion propagating forward in time interacting with a sink
pion propagating backwards in time and other similar contributions.  The constant
term was clearly evident in the $I=2$ channel, but was consistent with
zero in the $I=0$ channel.  Hence, the $I=0$ results shown in
Fig.~\ref{fig:twopionscorr} were done setting the constant term to zero.
This figure demonstrates that the stochastic LapH method can provide
sufficient accuracy to see the difference of these two-pion energies from the
energy of two free pions at rest, indicated by the horizontal dashed lines.
In the $I=1$ channel, the $\rho$-meson is expected to be the lowest-lying
energy level, so a correlator matrix including single-hadron and
two-pion operators is necessary to reliably extract the low-lying spectrum
in this channel.  This will be done in future work.  Again, we emphasize that
only very simple operators were used here, and
future use of better operators will improve the accuracy of these 
energies.

\section{Conclusion}
\label{sec:conclude}

A new method of stochastically estimating the low-lying effects of quark
propagation was proposed which allows accurate determinations of temporal
correlations of single-hadron and multi-hadron operators in lattice QCD.
The method enables accurate treatment of hadron correlators involving
quark propagation from all spatial sites on one time slice to all spatial
sites on another time slice.  Contributions involving quark lines 
originating at the sink time $t_F$ and terminating at the same sink time $t_F$
are easily handled, even for a large number of $t_F$ times.

The effectiveness of the method can be traced to two of its key features: the
use of noise dilution projectors that interlace in time and the use of
$Z_N$ noise in the subspace defined by the Laplacian Heaviside quark-field
smearing.  Introducing noise in the LapH subspace results in greatly
reduced variances in temporal correlations compared to methods that 
introduce noise on the entire lattice.  Although the number of Laplacian
eigenvectors needed to span the LapH subspace rises dramatically with the
spatial volume, we found that the number of inversions of the Dirac matrix
needed for a target accuracy was remarkably insensitive to the lattice volume,
once a sufficient number of dilution projectors were introduced.

In addition to increased efficiency, the stochastic LapH method has other
advantages.  The method leads to complete factorization of hadron 
sources and sinks in temporal correlations, which greatly simplifies 
the logistics of evaluating correlation matrices involving large numbers
of operators.  Implementing the Wick contractions of the quark lines is
also straightforward.  Contributions from different Wick orderings 
within a class of quark-line diagrams differ only by permutations of the
noises at the source. 

The method was tested using the isoscalar mesons in the scalar, pseudoscalar,
and vector channels, and using the two-pion system of total isospin $I=0,1,2$ 
on large anisotropic $24^3\times 128$ lattices with pion masses 
$m_\pi\approx 390$ and 240~MeV.  Given the success of these tests, we are now
applying the stochastic LapH method to compute the excitation spectrum of both
mesonic and baryonic stationary-states of QCD in large finite volume.

\begin{acknowledgments}
This work was supported by the U.S.~National Science Foundation 
under awards PHY-0510020, PHY-0653315, PHY-0704171, PHY-0969863, and
PHY-0970137, and 
through TeraGrid resources provided by the Pittsburgh Supercomputer Center, 
the Texas Advanced Computing Center, and the National Institute for Computational
Sciences under grant numbers TG-PHY100027 and TG-MCA075017.  MP is supported by 
Science Foundation Ireland under research grant 07/RFP/PHYF168. We 
acknowledge conversations with Balint Joo, Steve Wallace, David Richards,
Robert Edwards, Jozef Dudek, Christopher Thomas, and Huey-Wen Lin.
The USQCD QDP++/Chroma library\cite{Edwards:2004sx} was used in developing 
the software for the calculations reported here.  
\end{acknowledgments}

\bibliography{cited_refs}
\end{document}